\documentclass[aps,pra,twocolumn,footinbib]{revtex4-2}
\usepackage{graphicx}
\usepackage{indentfirst}
\usepackage{physics}
\usepackage{braket}
\usepackage{float}
\usepackage{amsmath}
\usepackage{epstopdf}
\usepackage{footnote}
\usepackage{CJK}
\usepackage{esint}
\usepackage{comment}
\usepackage{color}
\usepackage[T1]{fontenc}
\usepackage{subfigure}
\usepackage{amsfonts}
\usepackage{footmisc}
\usepackage{scrextend}
\usepackage{multirow}
\usepackage[hyperfootnotes=false]{hyperref}
\usepackage[acronym]{glossaries}
\usepackage[english]{babel}
\usepackage{url}
\usepackage{bm}
\usepackage{hyperref}
\usepackage{algpseudocode}
\usepackage{svg}
\definecolor{darkblue}{rgb}{0,0,0.5}
\hypersetup{
colorlinks=true,
linkcolor=black,
filecolor=blue,
citecolor=darkblue,  
urlcolor=black,
}

\newtheorem{theorem}{Theorem}
\newtheorem{algorithm}[theorem]{Algorithm}

\newcommand\argmin{\mathop{\mathrm{argmin}}}

\newcommand{\calC}{{\cal C}}

\newcommand{\calE}{{\cal E}}

\newcommand{\1}{^{(1)}}

\newcommand{\QZ}[1]{{{\textcolor{black}{#1}}}}

\renewcommand{\subsubsection}[1]{
\bigskip
{\noindent{\bf \normalsize#1.\ }}
}

\begin{document}


\title{Quantum Computational Phase Transition in Combinatorial Problems}
\author{Bingzhi Zhang$^{1,2}$}
\author{Akira Sone$^{3}$}
\author{Quntao Zhuang$^{1,4}$}
\email{zhuangquntao@email.arizona.edu}

\address{
$^1$Department of Electrical and Computer Engineering, University of Arizona, Tucson, Arizona 85721, USA
}
\address{
$^2$Department of Physics, University of Arizona, Tucson, AZ 85721, USA
}
\address{
$^3$Aliro Technologies, Inc, Boston, MA 02135, USA
}
\address{
$^4$James C. Wyant College of Optical Sciences, University of Arizona, Tucson, AZ 85721, USA
}
\begin{abstract}
Quantum Approximate Optimization algorithm (QAOA) aims to search for approximate solutions to discrete optimization problems with near-term quantum computers. As there are no algorithmic guarantee possible for QAOA to outperform classical computers, without a proof that $BQP\neq NP$, it is necessary to investigate the empirical advantages of QAOA. We identify a computational phase transition of QAOA when solving hard problems such as SAT---random instances are most difficult to train at a critical problem density.
We connect the transition to the controllability and the complexity of QAOA circuits. Moreover, we find that the critical problem density in general deviates from the SAT-UNSAT phase transition, where the hardest instances for classical algorithm lies.
Then, we show that the high problem density region, which limits QAOA's performance in hard optimization problems ({\it reachability deficits}), is actually a good place to utilize QAOA: its approximation ratio has a much slower decay with the problem density, compared to classical approximate algorithms. Indeed, it is exactly in this region that quantum advantages of QAOA over classical approximate algorithms can be identified.

\end{abstract}
\maketitle






\section{Introduction}
Quantum Approximate Optimization algorithm (QAOA)~\cite{farhi2014quantum}, like all quantum algorithms, aims to utilize quantum hardwares to efficiently solve problems that are hard on classical computers. It is one of the candidates to achieve a quantum supremacy in the noisy intermediate-scale quantum (NISQ) era~\cite{preskill2018quantum}. So far, quantum supremacy has only been realized for random circuit sampling tasks~\cite{arute2019quantum,wu2021strong}. For complex but practical problems in the class of nondeterministic-polynomial (NP) time, no quantum advantage has been found, despite trials of QAOA on the Google Sycamore quantum processor~\cite{harrigan2021quantum}. To search for quantum supremacy, it is crucial to first understand the difference between what is hard or easy for QAOA and for classical algorithms, which can be best explored via the combinatorial problem of Boolean satisfiability problem (SAT). 

In a $k$-SAT instance, one asks whether multiple clauses, each involving $k$ Boolean variables, can be satisfied simultaneously. Depending on the value of $k$, the worst-case hardness is drastically different---while 3-SAT is NP-complete, 2-SAT can be efficiently solved in polynomial time (class P). In addition, the classical empirical hardness of random 3-SAT instances is known to have a computational phase transition~\cite{cheeseman1991really,mitchell1992hard,achlioptas2001phase,leyton2014understanding,kalapala2005phase} versus the problem density characterized by the clause-to-variable ratio. When the density is small (large), almost all instances are satisfied (unsatisfied) and easy to solve; while for density approaching the critical point of the SAT-UNSAT phase transition, the 3-SAT problem instances become the hardest to solve. 

For quantum algorithms such as QAOA, the above phenomenon is largely unexplored. To begin with, as QAOA always implements SAT problems in their NP-hard optimization versions (Max-SAT)~\cite{haastad2001some}, it is unclear whether the decision version's NP ($k=3$) versus P ($k=2$) contrast has any influence on QAOA's performance. It is also unclear how classical empirical hardness of 3-SAT connects to QAOA's performance on its optimization versions. Indeed, Ref.~\cite{akshay2020reachability} does not find a big difference between QAOA's performance on Max-2-SAT and Max-3-SAT, and only finds QAOA's performance to worsen as the density increases---a phenomenon they call {\it reachability deficits}. 

In this paper, we reveal a computational phase transition in the trainability of QAOA in solving the positive $1$-in-$k$ SAT problem. 
In terms of trainability characterized by gradient, the typical amplitude of gradient in training \QZ{SAT problems achieves the minimum at a critical problem density ratio. In general, this quantum critical problem density deviates from the SAT-UNSAT phase transition~\cite{cheeseman1991really,mitchell1992hard,achlioptas2001phase,leyton2014understanding,kalapala2005phase}, where the instances are hardest classically.} 
We link this gradient transition to the controllability of the quantum systems evolving under the QAOA circuit~\cite{D'AlessandroBook08,Wang2016SubspaceControllability,D'Alessandro2010ConstructiveDecomposition,larocca2021diagnosing} and the complexity of QAOA circuit~\cite{dankert2009exact,roberts2017chaos,nahum2017quantum,zhuang2019scrambling}. In terms of accuracy of the optimization versions of SAT, we find the QAOA's approximation ratio to be robust and decay slowly with the problem density. Moreover, despite the performance decay due to reachability deficits, it is precisely in the large problem density region where a quantum advantage can be identified, \QZ{when comparing with classical approximate algorithms}. In addition, the accuracy in solving Max-2-SAT is higher than that in solving Max-3-SAT, consistent with the P versus NP contrast in the decision version of the problems. 
\QZ{Interestingly, for the decision version of the SAT problems, QAOA shows the worst performance at the SAT-UNSAT transition, revealing a remnant of classical empirical hardness. Such remnant of classical empirical hardness is also confirmed in quantum adiabatic algorithms (QAA)~\cite{farhi2001quantum, young2010first, zhuang2014increase}.}


\section{Results}

\subsection{Preliminary}
\subsubsection{Quantum Approximate Optimization algorithm}
To solve an optimization problem, QAOA encodes the cost function into the energy of a problem Hamiltonian $H_C$, defined over spin-1/2 particles (qubits), and then seeks for an approximation of the ground state that encodes the solution to the optimization problem. 
An $n$-qubit QAOA circuit implements dynamics governed by the problem Hamiltonian $H_{C}$ and a mixing Hamiltonian $H_B = \sum_{i=1}^n \sigma_i^x$ alternatively in each layer, where $\sigma_i^x$ is the Pauli-X operator representing the transverse fields. The output state of a $p$-layer QAOA is therefore
$
\ket{\psi(\vec{\gamma}, \vec{\beta})} = \prod_{\ell=1}^p e^{-i\beta_\ell H_B}e^{-i\gamma_\ell H_C}\ket{\psi\left(0, 0\right)},
$
where $\vec{\gamma}=\left(\gamma_1,\dots,\gamma_p\right)$ and $\vec{\beta}=\left(\beta_1,\dots,\beta_p\right)$ are variational parameters. The initial state is set to be a superposition of all possible spin configurations, $\ket{\psi\left(0, 0\right)} = \ket{+}^{\otimes n}$ with $\ket{+}=\left(\ket{0}+\ket{1}\right)/\sqrt{2}$. 
To solve the problem, variational training is performed over the parameters $\vec{\gamma}, \vec{\beta}$ to minimize the cost function 
$
\mathcal{C}(\vec{\gamma},\vec{\beta}) = \bra{\psi(\vec{\gamma},\vec{\beta})}H_C\ket{\psi(\vec{\gamma},\vec{\beta})}.
$
The variational training terminates when the cost function stops to decrease significantly, and ideally leads to the optimal parameters
$
\vec{\gamma}^*,\vec{\beta}^* = \argmin_{\vec{\gamma},\vec{\beta}}{\calC(\vec{\gamma},\vec{\beta})}.
$

\subsubsection{SAT problems}
\QZ{We will focus on two types of SAT problems, $k$-SAT problem ($k\ge 2$) and the positive $1$-in-$k$ SAT problem (1-$k$-SAT$^+$, $k\ge2$). The positive $1$-in-$k$ SAT problem (1-$k$-SAT$^+$, $k\ge2$) is also known as the exact-cover $k$ problem. Given $n$ Boolean variables $V = \{v_i\}_{i=1}^n$, a random instance of the SAT problems can be constructed by choosing $m$ clauses $C = \{c_a\}_{a=1}^m$, each containing $k$ different variables $\{v_{aj}\}_{j=1}^k$ uniformly randomly chosen from $V$. The $k$ elements in each clause can be either positive or negative literal in $k$-SAT problem with equal probability, while only positive in $1$-$k$-SAT$^+$ problems. The conjunctive normal form (CNF) of the SAT instance can be expressed as
$
F\left(V\right) = \bigwedge_{a=1}^m c_a\left(\{v_{aj}\}_{j=1}^k\right),
$
where `$\bigwedge$' denotes AND and forces the CNF to be true only when all clauses are satisfied.
In a $k$-SAT problem, each clause is true when at least one element in the clause is true; while in positive $1$-in-$k$ SAT, a clause $c_a\left(\{v_{aj}\}_{j=1}^k\right)$ is satisfied if and only if a single variable among $\{v_{aj}\}_{j=1}^k$ is taken to be true. }

\begin{figure}
    \centering
    \includegraphics[width=0.45\textwidth]{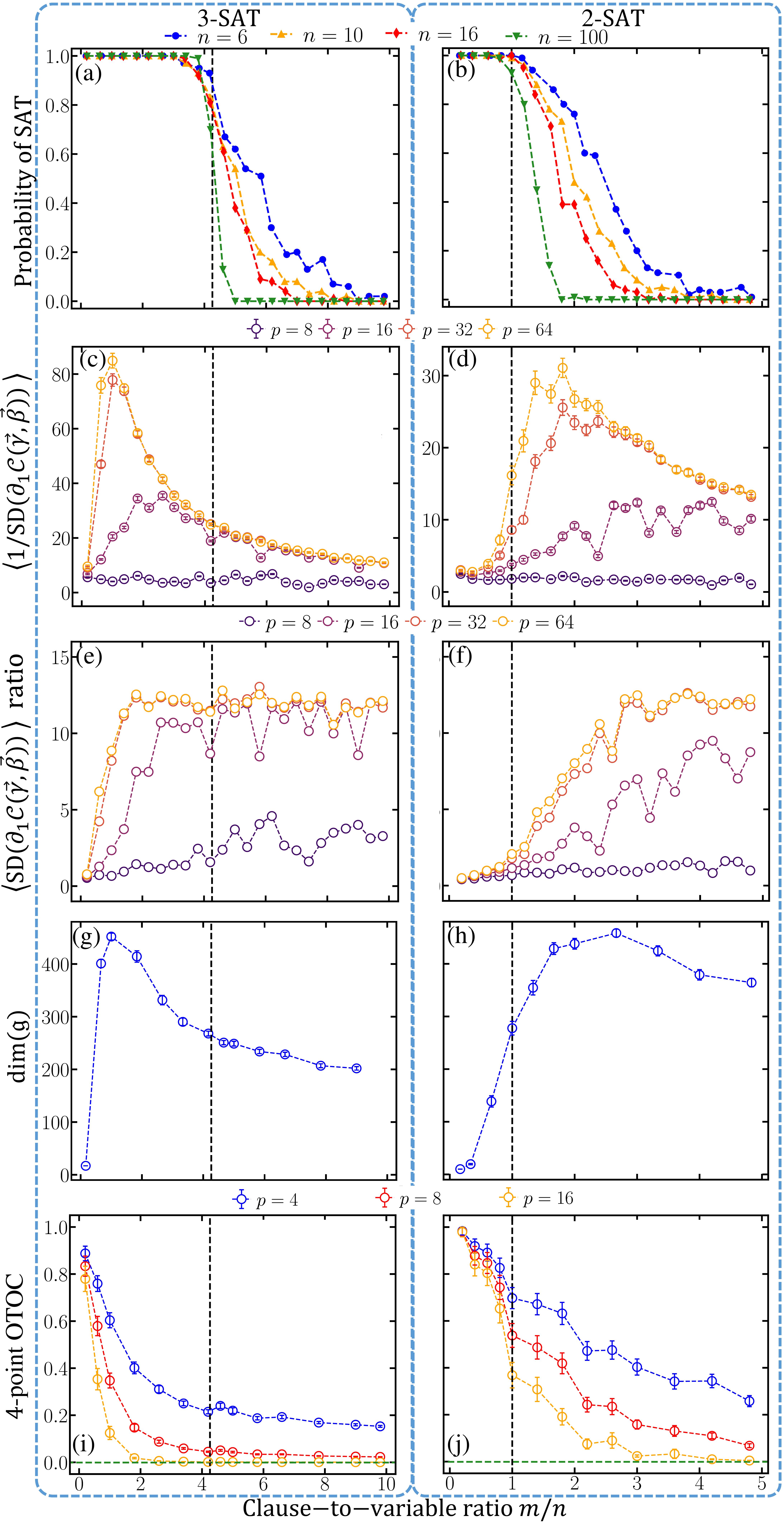}
    \caption{
    \QZ{
    {\bf SAT-UNSAT phase transition and trainability of $3$-SAT (left column) and $2$-SAT.}
    (a)(b) Probability of SAT for different system size $n$.
    (c)(d) The mean of $1/{\rm SD}\left(\partial_{1}\calC(\vec{\gamma},\vec{\beta})\right)$ in different $p$-layer QAOA with $n=16$ variables. The inverse is added for easier comparison according to Eq.~\eqref{eq:grad_DLA}.
    (e)(f) The ratio of average gradient variance $\braket{{\rm SD}(\partial_{1}\calC(\vec{\gamma},\vec{\beta}))}$ for $n=6$ variables over $n=16$ variables in different $p$-layer QAOA. Larger ratio indicates barren plateau.
    (g)(h) The dimension of dynamical Lie algebra ${\rm dim}(\mathfrak{g})$ for generators in QAOA in an $n=6$ qubits system.
    (i)(j) Average of $4$-point OTOC for different $p$-layer QAOA with $n=10$ variables. Green horizontal dashed line represents the value given by Haar unitary $-2^n/(4^n-1)$. 
    Vertical dashed lines indicate the critical SAT-UNSAT transition point. All results are evaluated over $100$ instances.
    }
    }
    \label{fig:pSAT_kSAT}
\end{figure}

\begin{figure}
    \centering
    \includegraphics[width=0.45\textwidth]{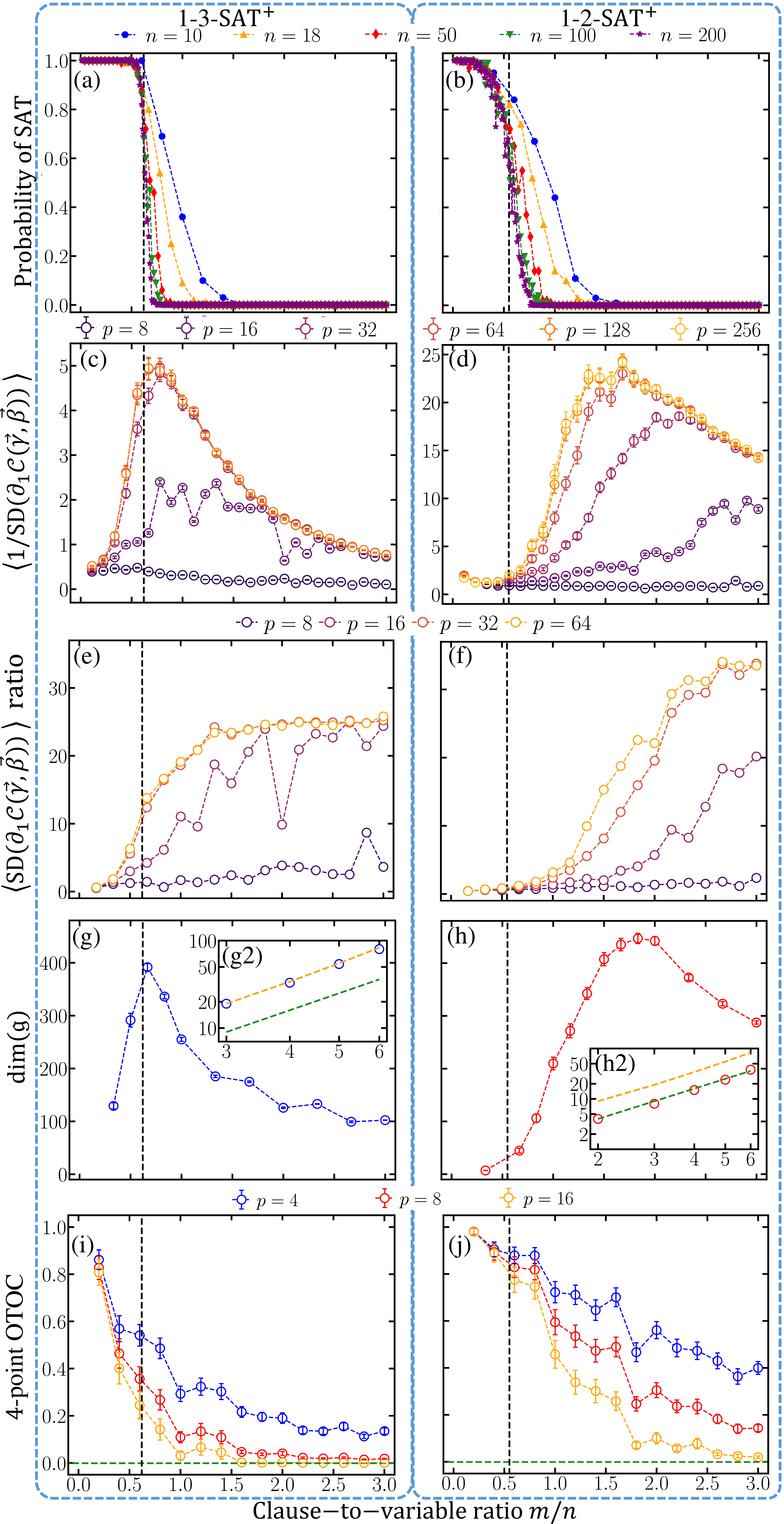}
    \caption{
    {\bf SAT-UNSAT phase transition and trainability of 1-$3$-SAT$^+$ (left column) and 1-$2$-SAT$^+$.}
    (a)(b) Probability of SAT for different system size $n$.
    (c)(d) The mean of $1/{\rm SD}\left(\partial_{1}\calC(\vec{\gamma},\vec{\beta})\right)$ in different $p$-layer QAOA with $n=18$ variables. \QZ{The inverse is added for easier comparison according to Eq.~\eqref{eq:grad_DLA}. (e)(f) The ratio of average gradient variance $\braket{{\rm SD}(\partial_{1}\calC(\vec{\gamma},\vec{\beta}))}$ for $n=6$ variables over $n=18$ variables in different $p$-layer QAOA. Larger ratio indicates barren plateau.}
    (g)(h) The dimension of dynamical Lie algebra ${\rm dim}(\mathfrak{g})$ for generators in QAOA in an $n=6$ qubits system. The inset log-log plots (g2) and (h2) show ${\rm dim}(\mathfrak{g})$ versus $n$ with a fully symmetric $H_C$. Green and orange curves represent the lower bound estimate ${\rm dim}(\mathfrak{g})=n^2$ and upper bound of $\dim^{\rm UB}$.
    (i)(j) Average of $4$-point OTOC for different $p$-layer QAOA with $n=10$ variables. Green horizontal dashed line represents the value given by Haar unitary $-2^n/(4^n-1)$. 
    Vertical dashed lines indicate the critical SAT-UNSAT transition point. All results are evaluated over $100$ instances.
    }
    \label{fig:pSAT}
\end{figure}

The (decision version of) SAT problem asks whether $F\left(V\right)$ can be satisfied with an assignment of variables $V$, while the optimization version---Max-SAT---aims to find an assignment of variables $V$ to minimize the number of clause violations. With the increase of clause-to-variable ratio $m/n$, it becomes harder to satisfy a random SAT instance, and there exists a phase transition of SAT probability across a critical ratio, $m/n = 1$ for $2$-SAT~\cite{goerdt1992threshold} and $m/n = 4.26$ for $3$-SAT~\cite{leyton2014understanding}, $m/n\sim 0.55$  for 1-$2$-SAT$^+$ and $m/n\sim0.62$ for 1-$3$-SAT$^+$~\cite{kalapala2005phase}, as shown in Fig.~\ref{fig:pSAT_kSAT}(a)(b) and Fig.~\ref{fig:pSAT}(a)(b).

We study the case of $k=2$ and $k=3$ for a comparison: while $2$-SAT and 1-$2$-SAT$^+$ are in class P and efficiently solvable, $3$-SAT and 1-$3$-SAT$^+$ are NP-complete and it takes an exponential amount of time to solve it, e.g., by the well-known algorithm X~\cite{knuth2000dancing}. Despite the contrast in the decision versions, Max-$k$-SAT and Max-1-$k$-SAT$^+$ are always NP hard, even for $k=2$~\cite{garey1974some}. In addition to the worst case hardness, empirical studies with classical algorithms on different variants of 3-SAT~\cite{cheeseman1991really,mitchell1992hard,achlioptas2001phase,leyton2014understanding,kalapala2005phase} show that when $m/n$ is small (large), almost all instances are satisfied (unsatisfied) and easy to solve; while for $m/n$ approaching the critical point of the SAT-UNSAT transition, the SAT problem instances become the hardest to solve. 

To solve SAT problems with QAOA, we transform each Boolean variables $v_i$ to the spin states of a qubit, with spin-down state $\ket{1}$ (Pauli-Z operator $\sigma^z=-1$) for true and spin-up state $\ket{0}$ ($\sigma^z=1$) for false, \QZ{and obtain the spin Hamiltonians for
$2$-SAT and $3$-SAT as
\begin{equation}
    H_{C,k} = \frac{1}{2^k}\sum_{a=1}^m \prod_{\ell=1}^k \left(1+A_{a_\ell, a}{\sigma^z_{a_\ell}}\right), \label{eq:ksat_H}
\end{equation}
where $A_{{a_\ell},a}$ stands for the literal sign for $\ell$th element in $a$th clause with $+1(-1)$ for positive(negative) literal separately and $0$ for absence of it in the clause.}
Similarly, for
1-$3$-SAT$^+$ and 1-$2$-SAT$^+$ as~\cite{farhi2001quantum, young2010first, zhuang2014increase, bengtsson2020improved} (see Methods)
\begin{subequations}
\begin{align}
    H_{C,3^+} &= \frac{1}{4}\sum_{a=1}^m \left(\sigma^z_{a1}+\sigma^z_{a2}+\sigma^z_{a3}-1\right)^2, \label{eq:p1in3sat_H}\\
    H_{C,2^+} &= \frac{1}{4}\sum_{a=1}^m \left(\sigma^z_{a1}+\sigma^z_{a2}\right)^2. \label{eq:p1in2sat_H}
\end{align}
\label{eq:p1inksat_H}
\end{subequations}
The gate-based implementation of the QAOA for our problem Hamiltonians can be found in Supplementary. Note~\ref{app:QAOA_implementation}. 
With the above encoding, an instance is satisfied only if the ground state energy is zero. As QAOA minimizes the cost function, it can be considered as an approximate algorithm for solving Max-1-$k$-SAT$^+$. By default, via a threshold decision, the solution of the optimization also implies the solution to the decision version. 


Our overall goal in this paper is to understand what is hard and what is easy on QAOA, both in terms of trainability and accuracy. 


\subsection{Gradient of QAOA training}
As a variational circuit, QAOA's cost-function gradients over variables $\vec{\gamma}, \vec{\beta}$ indicate the shape of cost-function landscape---larger amplitudes of gradients indicate sharper changes and therefore the problem is easier to train; while small amplitudes of gradients leads to barren plateaus~\cite{mcclean2018barren,cerezo2021cost,larocca2021diagnosing,larocca2021theory, larocca2021theory, wang2020noise} that make the training difficult. As gradients average to zero on random states~\cite{mcclean2018barren,cerezo2021cost}, we evaluate the standard deviation (SD) of gradients to characterize their typical amplitudes.

To represent the typical case of training, we evaluate the gradient on random choices of the circuit parameters via a numerical finite-difference. Without loss of generality, we consider the gradient over the first variable $\gamma_1$ and denote it as $\partial_1\calC$~\cite{mcclean2018barren,cerezo2021cost}.  To enable an easier visualization in Fig.~\ref{fig:pSAT_kSAT}(c)(d) and Fig.~\ref{fig:pSAT}(c)(d), we plot the inverse of the gradient SD, $1/{\rm SD}\left(\partial_1\calC(\vec{\gamma},\vec{\beta})\right)$, so that large values indicate hardness in convergence. We consider different number of layers $p$ in QAOA to obtain a comprehensive picture of it.

\QZ{For all the problems under study, the inverse gradient SD has a clear peak at a critical clause-to-variable $m/n$, as shown in Fig.~\ref{fig:pSAT_kSAT}(c)(d) and Fig.~\ref{fig:pSAT}(c)(d). However, this peak is in general different from the classical SAT-UNSAT transition indicated by the dashed line. For the special case of 1-$3$-SAT$^+$, Fig.~\ref{fig:pSAT}(c) shows that the peak of the inverse gradient coincides with the SAT-UNSAT transition.}
A large inverse gradient SD indicates a small gradient in the typical case, and therefore a more barren plateau that makes the training hard at the phase transition. When $p$ is small, the peak disappears; however, at the same time, QAOA fails to provide the accurate solution, making trainability irrelevant.

\QZ{We notice that the cases of $k=3$ have the inverse gradient peaked at a much smaller clause-to-variable density, as a result of the more complex clauses. Overall, the results reveal a transition of the trainability measured by gradient that is different from the classical SAT-UNSAT transition, showing that the empirical hardness for quantum algorithms can be different from classical algorithms.}


\subsubsection{Connection to controllability}
To understand the different behaviors of the gradient, we utilize the connection between gradient and controllability measured by the dimension of dynamical Lie algebra (DLA) of QAOA generators, as recently identified in Ref.~\cite{larocca2021diagnosing}. 

As explained in Ref.~\cite{D'AlessandroBook08}, DLA can be used to test the controllability of the quantum system governed by unitary dynamics. Let us consider an $n$-qubit system described by a Hilbert space $\mathcal{H}$.  Considering an optimal quantum control model described by a unitary $U=\prod_{k=1}^{K}e^{-iu_k H_k}\,$, where $\mathcal{G} \equiv \{H_1,\cdots,H_K\}$ is a set of generators and $\{u_1,\cdots,u_K\}\subseteq\mathbb{R}$ is a set of coefficients which are usually represented by the control fields. 
Then, the DLA $\mathfrak{g}\equiv \langle iH_1,\cdots,iH_K\rangle_{\rm Lie}\subseteq \mathfrak{su}(2^n)$,
is constructed by the repeated and nested commutators of the elements in $\mathcal{G}$.
The corresponding dynamical Lie group is therefore obtained by taking the exponential of the DLA $e^{\mathfrak{g}}\equiv\{e^{V_1}e^{V_2}\cdots e^{V_L},~V_{1},\cdots,V_{L}\in\mathfrak{g}\}.$
Generally, for a finite-time evolution governed by Schr\"{o}dinger's equation, the system is fully controllable when the set of the unitaries obtained during this evolution can cover all unitaries as its elements. This is precisely formulated by the so-called Lie algebra rank condition, which states that the system is fully controllable if and only if $\dim(\mathfrak{g})=4^n-1$~\footnote{Here, note that we suppose that  $\mathcal{G}$ does not include the identity without the loss of generality because the identity leads to the negligible global phase in QAOA scenario.}. For quantum systems where the whole Hilbert space is not fully controllable, when the DLA $\mathfrak{g}$ can be described as the direct sum, i.e. $\mathfrak{g}=\bigoplus_{j}\mathfrak{g}_j$, so that the Hilbert space can be written in a form of the direct sum of the subspace $\mathcal{H}_j$ as $\mathcal{H}=\bigoplus_{j}\mathcal{H}_j$, $\dim(\mathfrak{g}_j)$ determines the subspace controllability of $\mathcal{H}_j$~\cite{larocca2021diagnosing,Wang2016SubspaceControllability,D'Alessandro2010ConstructiveDecomposition}.

With a problem Hamiltonian $H_C$ and the mixing Hamiltonian $H_B$ as the generators, we can generate the DLA $\mathfrak{g}$ and provide an estimate of the standard deviation of gradient from the dimension of the DLA $\dim\left(\mathfrak{g}\right)$ as
\begin{equation}
    1/{\rm SD}\left(\partial_1\calC(\vec{\gamma},\vec{\beta})\right)\in \Omega\left( \left[{\rm poly}\left(\dim\left(\mathfrak{g}\right)\right)\right]^{1/2}\right)
    \label{eq:grad_DLA}
\end{equation}
where ``${\rm poly}$'' denotes a polynomial function. Here, for two functions $f(x)$ and $g(x)$, $f(x)\in\Omega(g(x))$ means $f(x)$ is bounded below by $g(x)$ asymptotically.

Therefore, we evaluate the DLA dimension numerically to compare with our gradient results. As the numerical evaluation is costly, we are limited to a smaller size of $n=6$. Despite the small size, as we see in \QZ{Fig.~\ref{fig:pSAT_kSAT}(g)(h) and Fig.~\ref{fig:pSAT}(g)(h)}, the DLA dimension $\dim\left(\mathfrak{g}\right)$ essentially has the same behavior versus the clause-to-variable ratio $m/n$, when compared with the inverse gradient; This manifests a clear connection between the gradient transition and the DLA dimension transition. 

\QZ{QAOA provides a clear physical insight to the concept of trainability of variational quantum algorithms on NISQ device. From Fig.~\ref{fig:pSAT_kSAT}(c)-(h) and Fig.~\ref{fig:pSAT}(c)-(h), the trainability and controllability have a trade off. This can be explained as the following. When the system is more (less) controllable, there are more (less) control protocols available to transform the initial state to the desired final state. Geometrically, these protocols can be described as the the accessible paths characterized by the parameters $(\vec{\gamma},\vec{\beta})$ from the initial state to the desired state. In this picture, our task is to find the optimal path from all the possible paths. Therefore, from the trainability perspective, it becomes harder (easier) to train $(\vec{\gamma},\vec{\beta})$ when the system is more (less) controllable. Despite being harder to train, the plurality of the paths also provides more hope to good performance. In addition, one can also connect DLA to controllability via the Quantum Fisher information matrix (QFIM)~\cite{larocca2021theory}. As the rank of QFIM characterizes the number of independent ways to vary the control parameters to change the generated quantum state, it is intuitive that the dimension of DLA upper bounds the rank of QFIM, which connects the training difficulty and controllability of a quantum model.}

An additional insight can be obtained by considering evaluating the DLA dimension at the $m\gg n$ limit for the 1-$k$-SAT$^+$ problems, where the Hamiltonian is symmetric between all qubits (see inset of Fig.~\ref{fig:pSAT} \QZ{(g)} and \QZ{(h)}). In this case,  we are able to prove an upper bound (see Methods), 
$
\dim(\mathfrak{g})\leq \dim^{\rm UB}\equiv\frac{1}{6}n(n^2+6n+11)\,.
$
We also expect $\dim(\mathfrak{g})$ to be above $n^2$, which is the dimension for a much simpler nearest neighbour Ising model~\cite{larocca2021diagnosing}.
While the upper bound is in general a loose one, it indicates that the gradient in the $m\gg n$ limit is only polynomially small; in contrast, for the hard instances we would expect an exponentially small gradient. This contrast supports the decay of dimension and the increase of gradient when $m/n$ is large.

\begin{figure}
    \centering
    \includegraphics[width=0.475\textwidth]{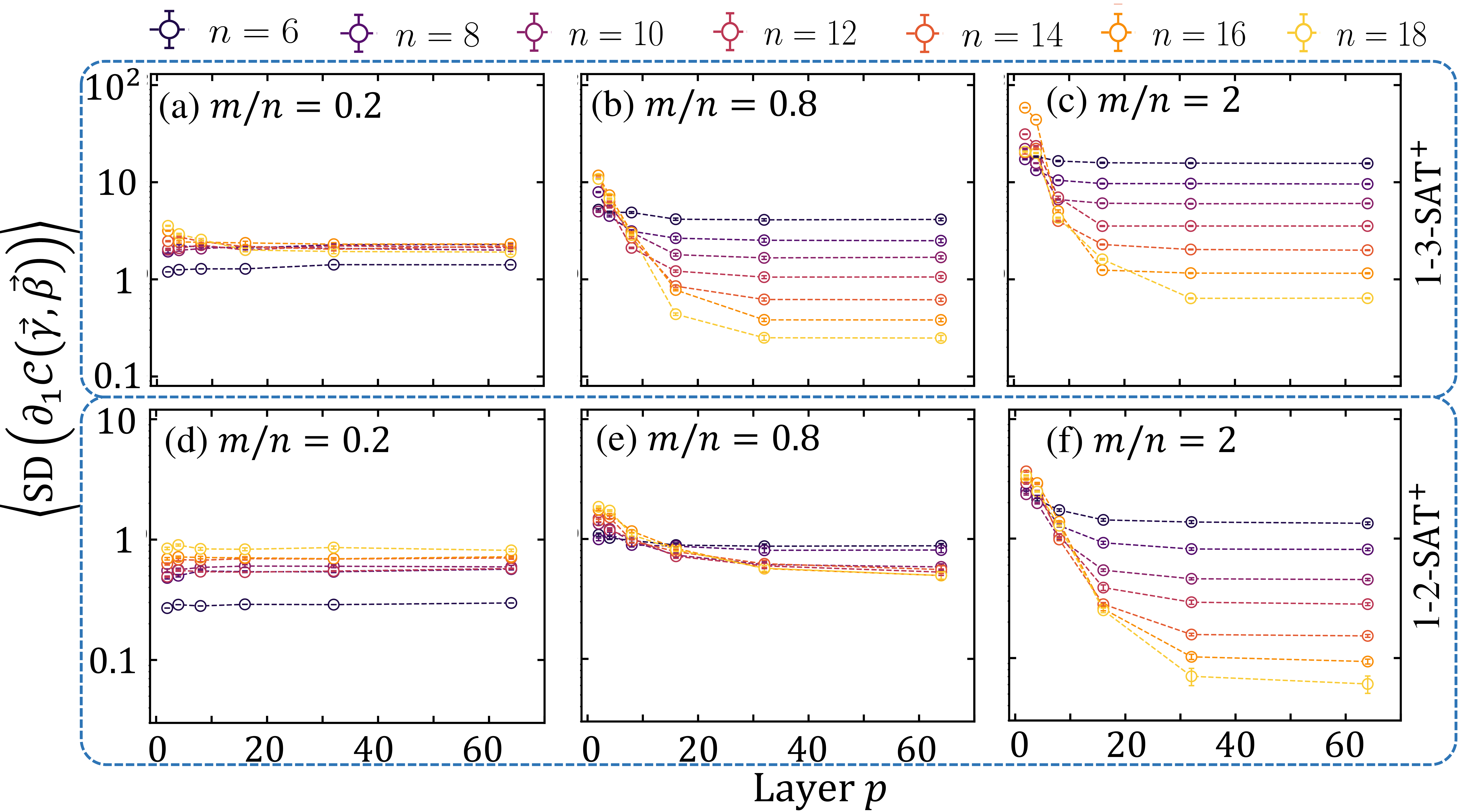}
    \caption{{\bf Barren plateau.} SD of gradient ${\rm SD}\left(\partial_{1}\calC\left(\gamma,\beta\right)\right)$ versus the layer of QAOA $p$ for 1-$3$-SAT$^+$ (top) for 1-$2$-SAT$^+$ (bottom) problems with different number of variables $n$. From left to right we plot at three different ratios $m/n=0.2, 0.8, 2$. Due to the finite size $n\le 18$, as shown in Fig.~\ref{fig:pSAT} the transition happens at around $m/n=0.8$.
    }
    \label{fig:gradient_vs_p}
\end{figure}

\subsubsection{Barren plateau and complexity}
\QZ{To further understand the barren plateau phenomena, we study the speed of decay of typical gradient with the number of qubits. To begin with, we pick two values of system size, $n=6$ and $n=16$ for $k$-SAT ($n=18$ for $1$-$k$-SAT$^+$ instead), and evaluate the ratio of the average gradient variance $\braket{{\rm SD}(\partial_{1}\calC(\vec{\gamma},\vec{\beta}))}$ for different number of layers $p$. In Fig.~\ref{fig:pSAT_kSAT} (e)(f) and Fig.~\ref{fig:pSAT} (e)(f), we see that right after the peak of the inverse gradients, when the circuit depth $p$ is sufficiently large, the decay ratio saturate to a value independent of the clause-to-variable ratio, indicating the barren plateau. While below the peak of the inverse gradient, the decay ratio of gradient increases gradually with the clause-to-variable ratio.}

To confirm an exponential decay of gradient, in Fig.~\ref{fig:gradient_vs_p}, we focus on 1-$k$-SAT$^+$ and plot the SD of gradient versus the layer $p$ for different number of qubits $n$, while keeping the clause-to-variable ratio $m/n$ to be a constant in each panel. For 1-$3$-SAT$^+$, when $m/n$ is small in panel (a), the SD of gradient saturates and does not decrease versus $p$ or $n$, showing no barren plateau; At the critical value in panel (b), we see an exponential decrease of SD versus the number of qubits $n$ at large $p$, confirming a barren plateau. Above threshold, as shown in panel (c), a barren plateau can still be confirmed, however, with larger gradients than the critical case of panel (b). On the contrary, for 1-$2$-SAT$^+$ as we see in panel (d) and (e), at around the SAT-UNSAT transition we do not see the appearance of a barren plateau. At large $m/n$ in panel (f), the gradient finally starts to show an exponential decay, indicating a barren plateau.

The appearance of barren plateaus are often connected to the complexity of the typical quantum circuit involved~\cite{cerezo2021cost}. The complexity of an ensemble of unitaries can in general be characterized by the closeness to unitary $t$-design, which reproduces the Haar random expectation values of $2t$-point correlators. In this regard, when the quantum circuit forms a $2$-design~\cite{dankert2009exact,roberts2017chaos,nahum2018operator}, it has been shown that the variance of the gradient will vanish exponentially with the system size---which leads to a barren \QZ{plateau} of cost function~\cite{mcclean2018barren,cerezo2021cost}.

Therefore, we consider the unitary ensemble $\mathcal{U}_p$ formed by the $p$-layer QAOA, $U_{\rm QAOA} = \prod_{\ell=1}^p e^{-i\beta_\ell H_B}e^{-i\gamma_\ell H_C}$, with each angle $\gamma_\ell\in [0,2\pi)$ and each angle $\beta_\ell \in [0,\pi)$ independent and uniform random. To measure the closeness to 2-design, we evaluate the ensemble-averaged infinite-temperature 4-point out-of-time-order correlator (OTOC) 
\begin{equation}
C_{\rm OTO}(W_1,W_2;\calE)=\frac{1}{d}\expval{ \Tr{W_1 U^\dagger W_2 U W_1 U^\dagger W_2 U}}_\calE,
\end{equation}
where the dimension $d=2^n$ for an $n$-qubit system and the average is over the unitary $U\in \calE$~\cite{roberts2017chaos}. For ensemble $\calE$ forming a 2-design, we have $C_{\rm OTO}(W_1,W_2;\calE)=-d/(d^2-1)$ saturate to the Haar results~\cite{nahum2018operator,roberts2017chaos}; while for trivial ensembles, $C_{\rm OTO}(W_1,W_2;\calE)$ is of order one. Therefore, the decay of OTOC indicates the ensemble being a 2-design. Without loss of generality, we consider the OTOC between single-qubit operators $C_{\rm OTO}(\sigma_1^y,\sigma_{n/2}^y;\mathcal{U}_p)$. \QZ{In Fig.~\ref{fig:pSAT_kSAT} (i)(j) and Fig.~\ref{fig:pSAT} (i)(j), we find the OTOC of the QAOA ensemble decays towards the Haar value when clause-to-variable ratio $m/n$ increases to the critical value of the minimum gradient, indicating a transition to 2-design. We also see a difference between the cases of $k=3$ versus $k=2$---the decay of OTOC for $k=2$ is much slower than $k=3$. 
}



\subsection{Accuracy of QAOA}
In this section, we explore the accuracy of QAOA in solving $k$-SAT and $1$-$k$-SAT$^+$. To speedup the training, we develop a heuristic pre-optimization initialization strategy (see Supplementary Note~\ref{app:QAOA_details}). To obtain the best accuracy, we perform $10$ repetitions on QAOA for each instance to obtain the optimal solution among those results. 
\QZ{To benchmark the accuracy of QAOA with the classical algorithm, in the case of Max-$k$-SAT, we consider the lower bound of state-of-the-art approximation algorithms; In the case of decision versions of $k$-SAT, we consider success probability of random guess; In the case of $1$-$k$-SAT$^+$, as less results are known about approximation ratios, we reduce the problem to the maximum weighted independent set (MWIS) problem~\cite{choi2010adiabatic,lucas2014ising} and utilize the greedy approximate MWIS algorithms proposed in~\cite{sakai2003note, kako2005approximation} (see Methods). }

The standard accuracy characterization of approximate algorithms for optimization problem is the approximation ratio~\cite{haastad2001some, sakai2003note, kako2005approximation}. For our case of Max-SAT problems, we define the approximation ratio $r\le 1$ of a solution to be the ratio between the number of clauses satisfied by the solution and the maximum number of clauses that can be satisfied by any solution. As the output state $\ket{\psi(\vec{\gamma},\vec{\beta})}$ in QAOA can be in a superposition of multiple solutions, we evaluate the expected approximation ratio via projecting the output state to the computational basis. \QZ{For Max-$k$-SAT, a random guess will satisfy on average $m(1-1/2^k)$ number of clauses; For Max 1-$k$-SAT$^+$, a random assignment will satisfy on average $mk/2^k$ number of clauses. For the instances with most clauses satisfiable, the above corresponds to an approximation ratio of $r_{\rm rand}\sim 1-1/2^k$ and $r_{\rm rand}\sim k/2^k$.} An exact optimal solution will saturate $r=1$ and non-trivial approximate algorithms should have $r\in[r_{\rm rand},1]$.

\begin{figure}[t]
    \centering
    \includegraphics[width=0.45\textwidth]{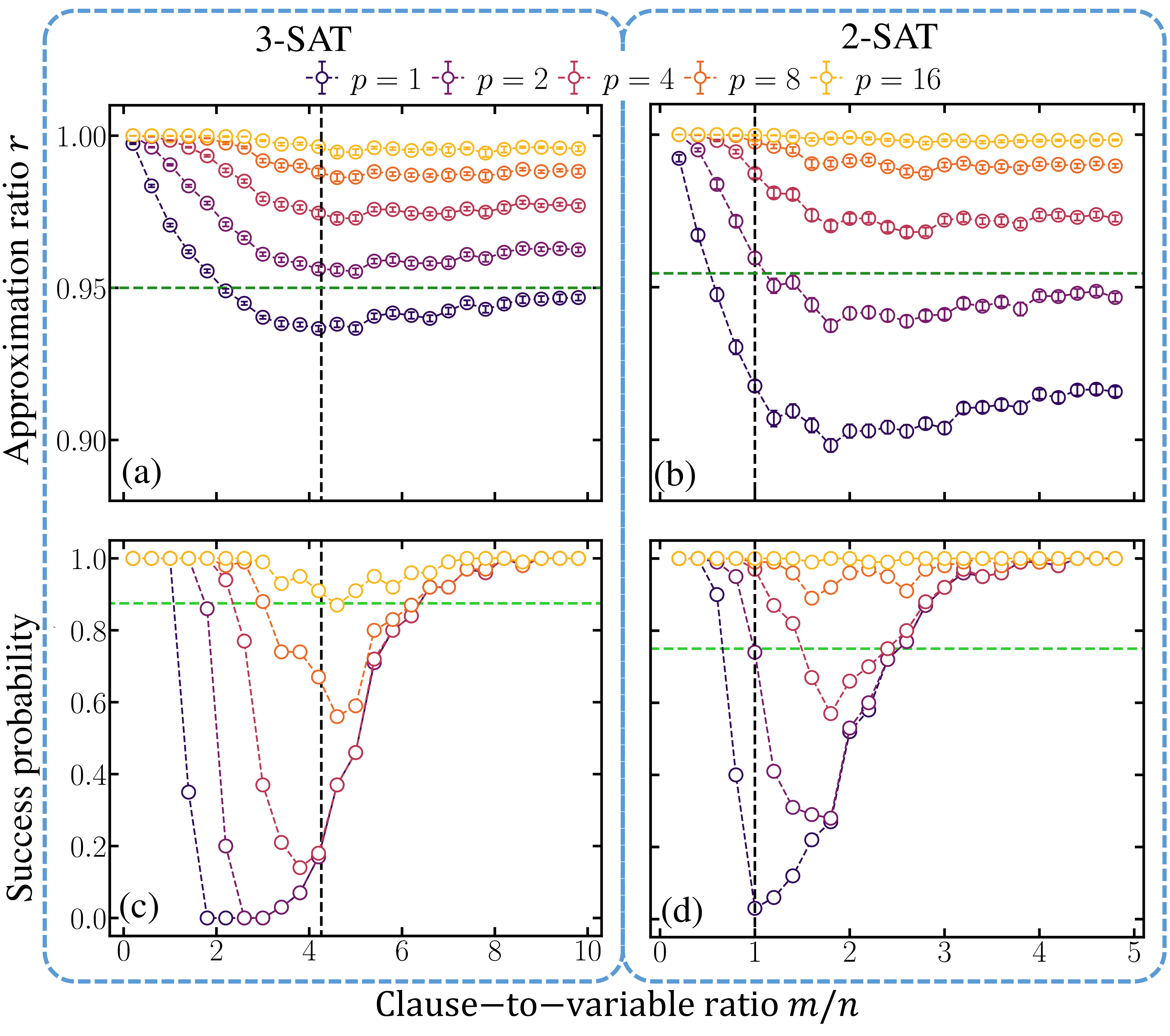}
    \caption{
    \QZ{{\bf Accuracy of QAOA.} (a)(b) Approximation ratio $r$ of SAT clauses, (c)(d) Success probability in determining SAT/UNSAT for $3$-SAT (left) and $2$-SAT (right) versus clause-to-variable ratio $m/n$ with $n=10$ variables. Green dashed line represent the lower bound of approximation algorithm $r\ge 0.95$ for Max-$3$-SAT~\cite{de20071} and $r\ge 21/22$ for Max-$2$-SAT~\cite{haastad2001some}. The horizontal light green dashed line in (c)(d) represent the success probability of the random guess which are $7/8$ and $3/4$ for $3$-SAT and $2$-SAT separately. Vertical black dashed lines in all plots represent critical point of SAT-UNSAT transition.}}
    \label{fig:qaoa_kSAT_ratio}
\end{figure}

\begin{figure}[t]
    \centering
    \includegraphics[width=0.45\textwidth]{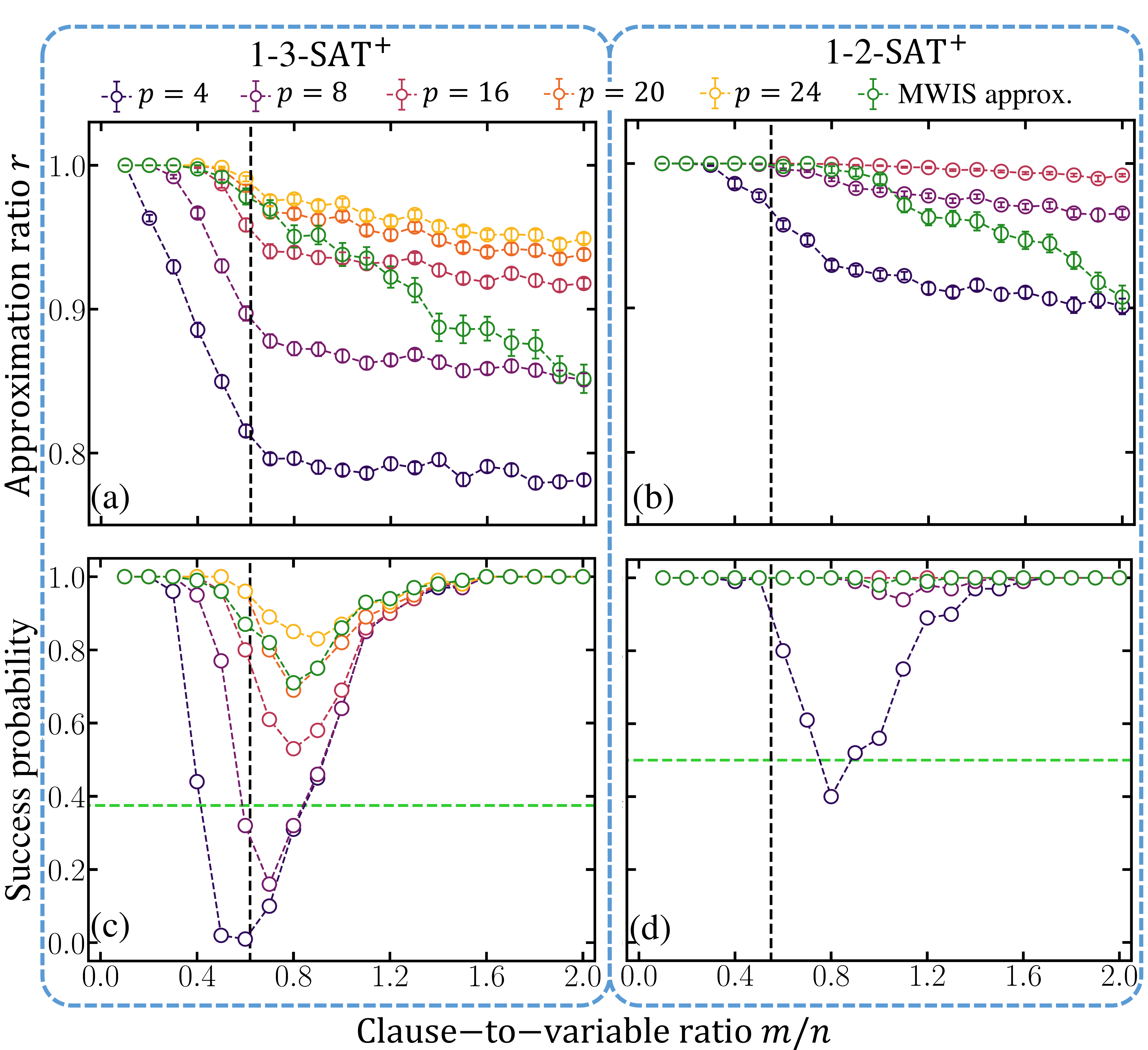}
    \caption{{\bf Accuracy of QAOA.} (a)(b) Approximation ratio $r$ of SAT clauses, (c)(d) Success probability in determining SAT/UNSAT for 1-$3$-SAT$^+$ (left) (from $p=4$ to $p=24$) and 1-$2$-SAT$^+$ (right) (from $p=4$ to $p=16$) versus clause-to-variable ratio $m/n$ with $n=10$ variables. Green dots represent the classical approximate results through a reduction to MWIS. The horizontal \QZ{light green} dashed line in (c)(d) represent the success probability of the random guess which are $3/8$ and $1/2$ for 1-$3$-SAT$^+$ and 1-$2$-SAT$^+$ separately. \QZ{Vertical black dashed lines in all plots represent critical point of SAT-UNSAT transition.}}
    \label{fig:qaoa_ratio}
\end{figure}

\QZ{In Fig.~\ref{fig:qaoa_kSAT_ratio} (a)(b) and Fig.~\ref{fig:qaoa_ratio} (a)(b), we see that as $p$ increases, QAOA is able to obtain larger approximation ratios. As the clause-to-variable ratio $m/n$ increases, the approximation ratio decays as expected. However, the decay is rather slow and manifest a robustness of QAOA. In the Max-$3$-SAT case, we see at small $p$ the approximation ratio is already better than the lower bound of $r\sim0.95$ in Ref.~\cite{de20071}, similarly, the lower bound of $r\ge 21/22$ for the Max-$2$-SAT~\cite{haastad2001some} case is also overcame at small depth $p$. In the case of Max 1-$k$-SAT$^+$, we consider the approximation ratio of the classical MWIS approximate algorithm for comparison (see Methods).} For Max-1-$3$-SAT$^+$, we identify a clear quantum advantage at around $p\sim 16$. For Max-1-$2$-SAT$^+$, advantages appear even for a shallow depth of $p=8$. We want to emphasize that the quantum advantage happens only when the clause-to-variable ratio is large, despite the reachability deficits~\cite{akshay2020reachability}. Indeed, we expect quantum algorithms to be advantageous especially for hard problems, where both classical and quantum algorithms face challenges. 


\QZ{Although all Max-SAT problems being considered are NP-hard, we do see some interesting contrast in the performance. In the Max-$2$-SAT and Max-$3$-SAT cases, the approximation ratio performance is similar when $p$ is large, consistent with previous results in Ref.~\cite{akshay2020reachability}; however, for the absolute number of additional violated clauses, Max-$2$-SAT performs slightly better than Max-$3$-SAT (see Supplementary Note~\ref{app:QAOA_details}). In the Max-1-$k$-SAT$^+$ cases, the accuracy of QAOA is substantially higher for $k=2$ than $k=3$ with the same number of $p$ layers. We speculate such a contrast in the performance can be caused by the different connectivity and complexity of the Hamiltonian in the problems.}


To connect to the empirical hardness transition in classical \QZ{algorithms}, we can also reinterpret each optimization result as a decision of SAT/UNSAT. This can be done via a threshold decision on the minimized number of UNSAT clauses, {\it e.g.}, determine an instance as SAT when the expected number of UNSAT clauses is smaller than $E_{\rm th}=0.5$ and UNSAT otherwise. To characterize the overall performance, we evaluate the success probability of deciding SAT/UNSAT when solving random instances at a fixed clause-to-variable ratio $m/n$. The results are shown in Fig.~\ref{fig:qaoa_kSAT_ratio} (c)(d) and Fig.~\ref{fig:qaoa_ratio}(c)(d).

The success probability increases with the layer of QAOA $p$ as we expect. For the $k=3$ cases in Fig.~\ref{fig:qaoa_kSAT_ratio} (c) and Fig.~\ref{fig:qaoa_ratio}(c), there is a valley of low success probability at around the critical point of $m/n$ shown in Fig.~\ref{fig:pSAT}(a), recovering the same hardness transition identified in empirical studies of classical algorithms~\cite{cheeseman1991really,mitchell1992hard,achlioptas2001phase,leyton2014understanding,kalapala2005phase}. While for the $k=2$ cases in Fig.~\ref{fig:qaoa_kSAT_ratio} (d) and Fig.~\ref{fig:qaoa_ratio}(d), despite a similar valley of low success probability at small $p$, the success probability is almost unity for a circuit depth of $p=16$. Similarly, the classical benchmark can be reinterpreted and similar transition versus $m/n$ can be seen. \QZ{Such a valley at the classical SAT-UNSAT transition indicates a remnant of the classical empirical hardness and is different from the trainability transition identified in Fig.~\ref{fig:pSAT_kSAT} and Fig.~\ref{fig:pSAT}.}

Combining the above, we see that overall QAOA possesses a similar notion of what is hard and easy as classical algorithms, while showing advantage over \QZ{the} classical algorithms \QZ{being considered} in the large problem-density instances. 

\begin{figure}[t]
    \centering
    \includegraphics[width=0.45\textwidth]{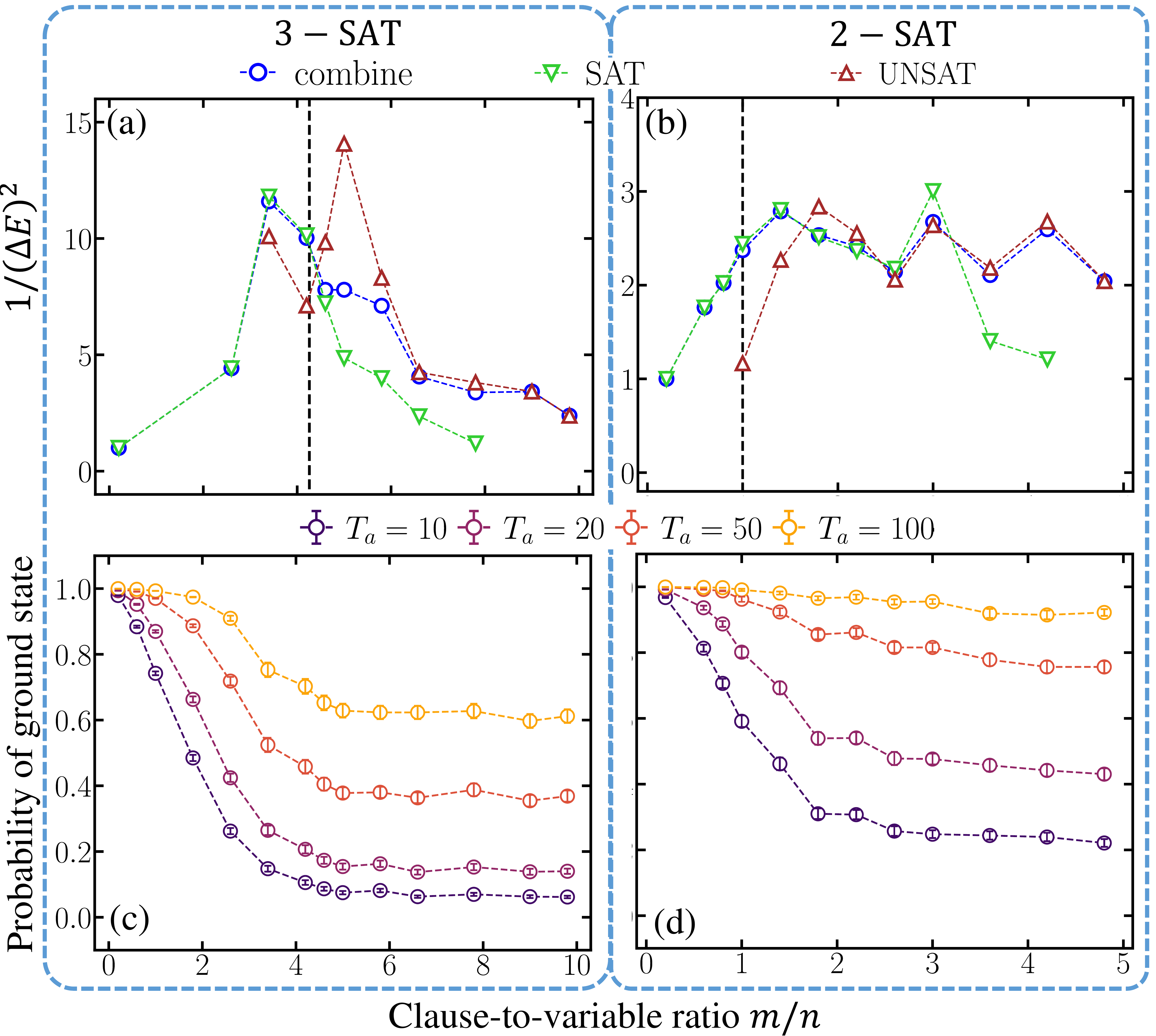}
    \caption{\QZ{{\bf QAA gap size and performance for $k$-SAT problems.}(a)(b) The median of $1/\Delta E^2$ of QAA with $n=10$ variables shown by blue circles. The green and purple circles represent the SAT and UNSAT instances separately. (c)(d) The probability that state through QAA evolution lies in the ground state $P=\sum_{i=1}^D|\braket{\psi_i|\phi_{\rm QAA}}|^2$.}}
    \label{fig:QAA_ksat}
\end{figure}

\begin{figure}[t]
    \centering
    \includegraphics[width=0.45\textwidth]{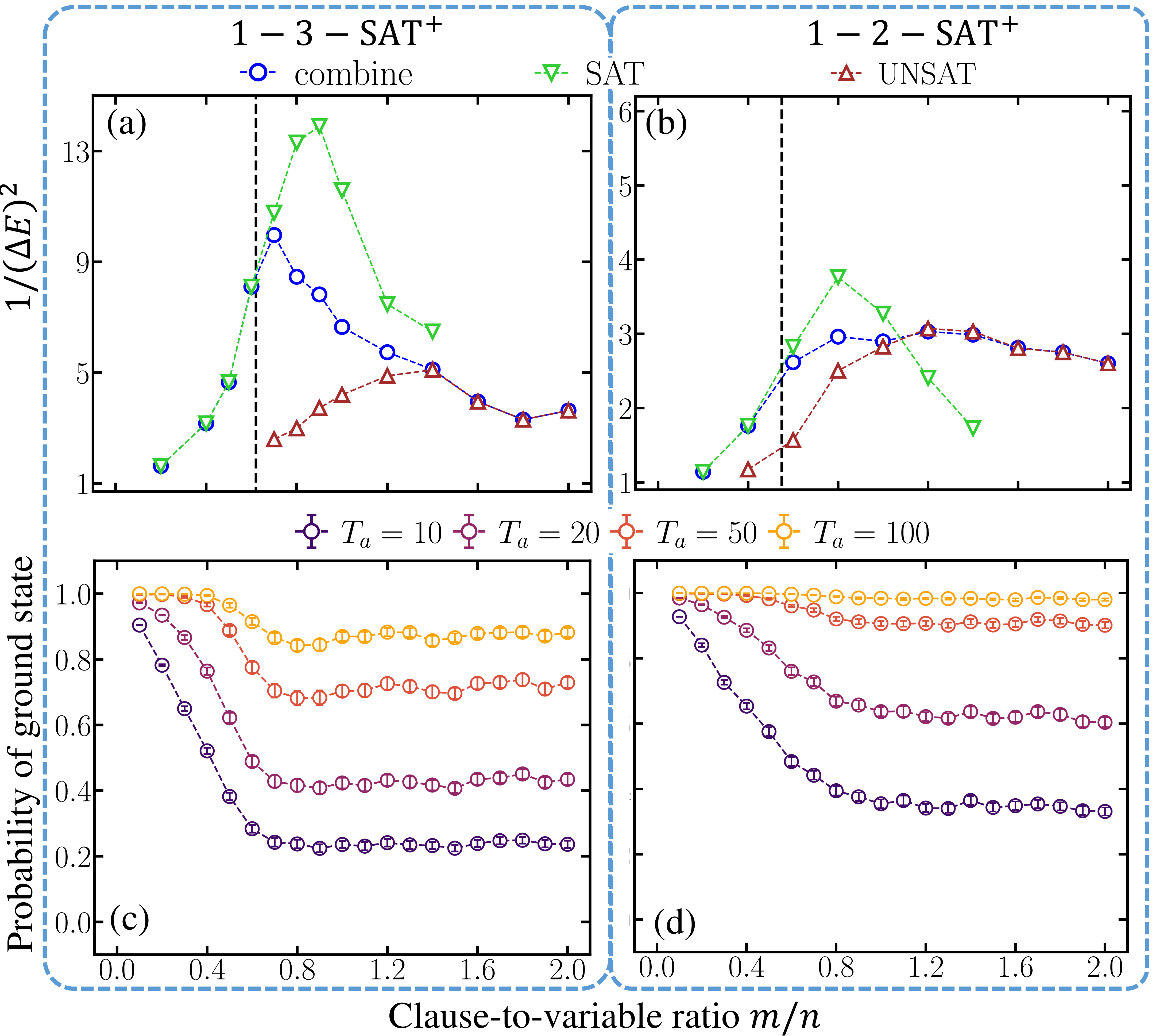}
    \caption{{\bf QAA gap size and performance of $1$-$k$-SAT$^+$.}(a)(b) The median of $1/\Delta E^2$ of QAA with $n=10$ variables shown by blue circles. The green and purple circles represent the SAT and UNSAT instances separately. (c)(d) The probability that the state through QAA evolution lies in ground state $P=\sum_{i=1}^D |\braket{\psi_i|\phi_{\rm QAA}}|^2$.}
    \label{fig:QAA}
\end{figure}

\subsection{Comparison with quantum adiabatic algorithm}

\QZ{The identified trainability transition in general deviates from the classical computational phase transition, while the performance in solving the SAT problems show a consistent trend with the classical computational phase transition. Such a disparity intrigues us to explore the empirical hardness of SAT instances in other Hamiltonian-based quantum algorithms, such as QAA---a popular alternative and also the predecessor of QAOA.}

To obtain the solution, QAA prepares the ground state of the problem Hamiltonian in Eqs~\eqref{eq:ksat_H} and~\eqref{eq:p1inksat_H} via an adiabatic evolution of the Hamiltonian 
\begin{equation}
H(s)=s H_C+(1-s)H_B^\prime, s\in[0,1],
\end{equation} 
from an ancillary Hamiltonian $H_B^\prime$ at $s=0$ to the problem Hamiltonian $H_C$ at $s=1$. Here the ancillary Hamiltonian $H_B^\prime=\sum_{i=1}^n |h_i|\sigma_i^x$, where $|h_i|$ is the number of times the variable $v_i$ appear in the clauses (see Methods). The initial Hamiltonian at $H(0)=H_B^\prime$, with a easy to prepare ground state $\ket{\psi\left(0\right)} \propto \left(\ket{0}-\ket{1}\right)^{\otimes n}$ in a superposition of all possible spin configurations. As one tunes the parameter $s$ slowly towards $H(1)=H_C$, the adiabatic theorem guarantees that the state of the system stays in the ground state; therefore, the final state $\ket{\phi_{\rm QAA}}$ is the ground state of the problem Hamiltonian, which provides the solution to the optimization problem.

From the adiabatic theorem, we can obtain an estimation on the computation time of QAA as $T_a \sim 1/(\Delta E)^2$ so that the success probability is close to unity, where $\Delta E$ is the minimum gap of the Hamiltonians $\{H(s), s\in [0,1]\}$~\cite{farhi2001quantum,zhuang2014increase}. In Fig.~\ref{fig:QAA_ksat}(a)(b) and Fig.~\ref{fig:QAA}(a)(b), we evaluate the inverse gap square $1/(\Delta E)^2$ as a measure of the instance hardness with different clause-to-variable ratio using Qutip~\cite{johansson2012qutip}. \QZ{Note that there exist other more rigorous estimations on the necessary adiabatic evolution time, combining higher-order terms~\cite{cheung2011improved,albash2018adiabatic}. However, the inverse gap square as a approximate estimation is sufficient for our purpose. In subplots (a), we identify a computational phase transition for $3$-SAT and 1-$3$-SAT$^+$, where the minimum gap is minimum at about the critical SAT/UNSAT transition, up to some small deviation due to finite size, similar to the decision version of SAT on QAOA in subplots (c) of Fig.~\ref{fig:qaoa_kSAT_ratio} and Fig.~\ref{fig:qaoa_ratio}.} While for $2$-SAT and 1-$2$-SAT$^+$, we see the minimum gap to be higher than the critical point, qualitatively agreeing with the case of QAOA. 

To further confirm the transition, we evaluate the success probability $P=\sum_{i=1}^D |\braket{\psi_i|\phi_{\rm QAA}}|^2$ from the overlap between the final evolved state $\ket{\phi_{\rm QAA}}$ and the $D$-degenerate ground state of the problem Hamiltonian $\ket{\psi_i}$. \QZ{In Fig.~\ref{fig:QAA_ksat}(c)(d) and Fig.~\ref{fig:QAA}(c)(d), at a finite time $T_a$, we see the success probability of QAA decreases before the critical point, while roughly maintaining a constant above the critical point. Such a robustness to the problem density coincides with the slow decay of QAOA's approximation ratio with the problem density. In addition, we also find the success probability of $2$-SAT and 1-$2$-SAT$^+$ (subplots (d)) to be much higher than that of $3$-SAT and 1-$3$-SAT$^+$ (subplots (c)).}

\section{Discussion}
\QZ{In this paper, we thoroughly explore the empirical hardness of Hamiltonian-based quantum algorithms in solving SAT problems. In the case of QAOA, we find a trainability phase transition, where the the gradient is minimum at certain critical problem density.} 
Such a phase transition is connected to the controllability and complexity of QAOA circuits. \QZ{Although the trainability transition in general deviates from the classical SAT-UNSAT transition, in terms of performance, Hamiltonian-based algorithms do show a remnant of the classical SAT-UNSAT transition.} Although our results are empirical, we expect analytical results to be challenging, as the classical correspondence of such transition is also empirical due to the complexity of the SAT problems.
We also identify quantum advantages of QAOA against \QZ{several} classical greedy approximate algorithms for a relatively small-scale quantum system, potentially realizable in the near-term.
Although we have focused on two cases of SAT for convenience, we expect the computational phase transition in QAOA to apply to all combinatorial optimization problems. In particular, as 3-SAT is NP-complete, the clause-to-variable ratio represents a universal characterization of a `problem density'~\cite{akshay2020reachability}, and the computational phase transition applies to all NP-complete problems in this regard.



\section{Methods}

\subsection{Many-body Formulation of the problem Hamiltonian} 
Here we introduce the many-body formulation of the problem spin Hamiltonian in Eqs.~\eqref{eq:ksat_H} and \eqref{eq:p1inksat_H}. For convenience, we first introduce an $n\times m$ binary matrix $A_{ij}$, where $A_{ij}=1(-1)$ if the variable $v_i$ is included in the clause $c_j$ as positive(negative) literal and $A_{ij}=0$ otherwise; With this matrix in hand, we can express all Hamiltonians in a standard many-body form~\cite{lucas2014ising} as
\begin{subequations}
\begin{align}
    H_{C,3} &= \frac{1}{8}\sum_{i<j<\ell} K_{ij\ell} \sigma_{i}^z\sigma_{j}^z\sigma_{\ell}^z + \frac{1}{8}\sum_{i<j} J_{ij} \sigma_{i}^z\sigma_{j}^z - \frac{1}{8}\sum_i h_i \sigma_{i}^z + \frac{m}{8} \label{eq:3sat_H_body}\\
    H_{C,2} &= \frac{1}{4}\sum_{i<j} J_{ij} \sigma_{i}^z\sigma_{j}^z - \frac{1}{4}\sum_i h_i \sigma_{i}^z + \frac{m}{4} \label{eq:2sat_H_body}
\end{align}
\end{subequations}
for $k$-SAT problems; and
\begin{subequations}
\begin{align}
    H_{C,3^+} &= \frac{1}{2}\sum_{i=1}^n h_i \sigma_i^z + \frac{1}{2}\sum_{i<j}J_{ij}\sigma_i^z\sigma_j^z + m, \label{eq: p1in3sat_H_ising}\\
    H_{C,2^+} &= \frac{1}{2}\sum_{i<j} J_{ij}\sigma_i^z\sigma_j^z + \frac{m}{2}, \label{eq:p1in2sat_H_ising}
\end{align}
\label{eq:H_ising}
\end{subequations}
for $1$-$k$-SAT$^+$ problems. The notations are introduced as 
\begin{subequations}
\begin{align}
    h_i &= -\sum_{j=1}^m A_{ij}\\
    J_{ij} &= \sum_{a=1}^m A_{ia}A_{ja}\\
    K_{ij\ell} &= \sum_{a=1}^m A_{ia}A_{ja}A_{\ell a}
\end{align}
\end{subequations}
Note that the number of clauses containing $v_i$ is equal to $|h_i|$.

\subsection{Dynamical Lie algebra: definition and bounds} 
\label{app:DLA}

Below we give bounds for $\dim\left(\mathfrak{g}\right)$ in the $m\gg n$ limit for both 1-$2$-SAT$^+$ and 1-$3$-SAT$^+$, where all coefficients $J_{ij}$'s and $h_i$'s approach uniform (see Supplementary Note~\ref{Hamiltonian_coefficients}).

For 1-$2$-SAT$^+$, we have 
$
H_{C,2^+} \propto  \sum_{i<j}\sigma_i^z\sigma_j^z
$
up to a constant.
Then, the set of the initial generators for the corresponding DLA $\mathfrak{g}_{H_{C,2^+},H_B}$ is
$
\mathcal{G}_{2^+}\equiv\left\{\sum_{i=1}^{n}\sigma_i^x,~\sum_{i<j}\sigma_i^z\sigma_j^z\right\}\,.
$
For the fully coupled Ising model with transverse fields along $x$ and $y$ axis, the set of the initial generators becomes 
$\mathcal{G}_{x,y}\equiv\left\{\mathcal{G}_2, \sum_{i=1}^{n}\sigma_i^y\right\}$.
From Ref.~\cite{Albertini08}, the dimension of the corresponding DLA $\mathfrak{g}_{x,y}$ is
\begin{equation}
    \dim(\mathfrak{g}_{x,y}) = \binom{n+3}{n}-1=\frac{1}{6}n(n^2+6n+11)\,,
\end{equation}
where $\binom{a}{b}\equiv a!/(a-b)!b!$ is the binomial coefficient.
Since the DLAs are generated by the repeated and nested commutators of the generator sets, we must have $\dim(\mathfrak{g}_{H_{C,2^+},H_B})\leq\dim(\mathfrak{g}_{x,y})$ due to $\mathcal{G}_2\subset \mathcal{G}_{x,y}$, which leads to 
\begin{equation}
     \dim(\mathfrak{g}_{H_{C,2^+},H_B})\leq \frac{1}{6}n(n^2+6n+11)\,.
\end{equation} 
Also we know nearest neighbour Ising model has dimension $n^2$~\cite{larocca2021diagnosing}. Therefore, we expect the scaling to be between $\Omega(n^2)$ and $O(n^3)$.

For the 1-$3$-SAT$^+$, we have 
$H_{C,3^+} \propto \frac{2}{n-1}\sum_{i<j} \sigma_i^z\sigma_j^z -\sum_{i=1}^n \sigma_i^z$. Then, the initial set of generators is $\mathcal{G}_3=\left\{\sum_{i=1}^{n}\sigma_i^x,~\frac{2}{n-1}\sum_{i<j} \sigma_i^z\sigma_j^z -\sum_{i=1}^n \sigma_i^z\right\}$. Let $\mathfrak{g}_{H_{C,3^+},H_B}$ be the corresponding DLA. Here, because we can write
\begin{align}
    e^{i H_{C,3^+}} = e^{i\frac{2}{n-1}\sum_{i<j}\sigma_i^z\sigma_j^z} e^{-i\sum_{i=1}^n\sigma_i^z}\,,
\end{align}
if we start from the initial set of generator
$\mathcal{G}_3^\prime=
\left\{\sum_{i=1}^{n}\sigma_i^x,
~\sum_{i=1}^n \sigma_i^z,
~\sum_{i<j}\sigma_i^z\sigma_j^z
\right\}$, we have the corresponding DLA to strictly contain $\mathfrak{g}_{H_{C,3^+},H_B}$. Now, due to the commutator $[\sum_{i=1}^{n}\sigma_i^x, \sum_{i=1}^{n}\sigma_i^y]\propto \sum_{i=1}^{n}\sigma_i^z$, the corresponding DLA of $\mathcal{G}_3^\prime$ becomes exactly $\mathfrak{g}_{x,y}$. Therefore, we have ${\rm dim}(\mathfrak{g}_{H_{C,3^+},H_B}) \le {\rm dim}(\mathfrak{g}_{x,y})$, which leads to
\begin{equation}
     \dim(\mathfrak{g}_{H_{C,3^+},H_B})\leq \frac{1}{6}n(n^2+6n+11)\,.
\end{equation}
The lower bound estimation $n^2$ for the dimension of DLA from nearest neighbour Ising model still holds.

\subsection{Classical approximate algorithms}
To solve Max-1-$k$-SAT$^+$, we transform it to the MWIS problem. Given a 1-$k$-SAT$^+$ instance with $n$ variables and $m$ clauses, one can construct a weighted graph with $n$ vertices $\{q_i\}_{i=1}^n$, each corresponding to a variable $v_i$ and having the weight $w(q_i)=|h_i|\ge 0$. For every two distinct vertices $q_i, q_j$, an edge $(q_i, q_j)$ exists if $J_{ij}>0$---when the corresponding variable $v_i, v_j$ appear in at least one clause at the same time. 

One can verify that the SAT/UNSAT version of 1-$k$-SAT$^+$ problem is reduced to asking whether the weight of maximum independent set is equal to $m$ or not. The reason is simple: an independent set of this graph corresponds to an assignment that does not have more than one true assignment in any clause. To guarantee a solution to the 1-$k$-SAT$^+$ instance, we still need to make sure that all clauses have one true variable. As the total weight of the independent set is equal to how many clauses are satisfied by this assignment, therefore if the total weight is equal to $m$, all clauses are satisfied. At the same time, the Max-1-$k$-SAT$^+$ can be reduced to solving the MWIS. As a classical benchmark, we can utilize various greedy algorithms for MWIS~\cite{sakai2003note, kako2005approximation} and choose the best performance among them (see Supplementary Note~\ref{app:MWIS-algorithms}).

\bigskip
\noindent{\bf \normalsize ACKNOWLEDGEMENTS}\\
\noindent
This work is supported by Defense Advanced Research Projects Agency (DARPA) under Young Faculty Award (YFA) Grant No. N660012014029, U.S. Department of Energy, Office of Science, National Quantum Information Science Research Centers, Superconducting Quantum Materials and Systems Center (SQMS) under the contract No. DE-AC02-07CH11359, National Science Foundation (NSF) Engineering Research Center for Quantum Networks Grant No. 1941583 and National Science Foundation (NSF) CAREER Award CCF-2142882. A.S. is supported by the internal R\&D from Aliro Technologies, Inc. Q.Z. and A.S. acknowledge helpful discussions with Marco Cerezo and Francesca Albertini.

\bigskip
\noindent{\bf \normalsize DATA AVAILABILITY}\\
\noindent
The data that support the findings of this study are available upon reasonable request.

\bigskip
\noindent{\bf \normalsize CODE AVAILABILITY}\\
\noindent
The code used to generate data will be made available to the interested reader upon reasonable request.

\bigskip
\noindent{\bf \normalsize COMPETING INTERESTS}\\
\noindent
The author declares no competing interests.

\bigskip
\noindent{\bf \normalsize AUTHOR CONTRIBUTIONS}\\
\noindent
Q.Z. proposed the study during a discussion with A.S. B.Z. performed the numerical calculations, analyzed the data and generated the figures, under the supervision of Q.Z. 
B.Z. and Q.Z. wrote the initial version of the manuscript. A.S. pointed out the connection to controllability, proved the bounds on DLA dimension and wrote the corresponding paragraphs, with inputs from Q.Z. and B.Z.  
All authors contributed to the writing of the final version of the manuscript.

\section{Supplementary Notes}

\subsection{Distribution of Hamiltonian coefficients}
\label{Hamiltonian_coefficients}

In this section, we analytically derive the distribution of coefficients in the problem Hamiltonian $H_{C,k^+}$ for both $k=3$ and $k=2$ (see Eqs.~\eqref{eq:H_ising}), and as well as the mean and variance. 

For 1-$3$-SAT$^+$, the probability that $A_{ia}A_{ja} = 1$ for arbitrary two different $i$ and $j$ is $p_3 = 3/\binom{n}{2}$. According to the definition $J_{ij} = \sum_{a=1}^m A_{ia}A_{ja}$, the probability that $J_{ij} = J$ is
\begin{equation}
    P_3\left(J_{ij}=J\right) = \binom{m}{J}p_3^J\left(1-p_3\right)^{m-J},
    \label{eq:Jdist_3}
\end{equation}
where we can directly see that the distribution of $J$ is dependent on the number of variables $n$ and number of clauses $m$. With the distribution of $J_{ij}$, the mean and variance are
\begin{subequations}
\begin{align}
    \mathbb{E}_3\left(J_{ij}\right) &= \frac{6m}{n\left(n-1\right)},\\
    {\rm Var}_3\left(J_{ij}\right) &= \frac{6m\left(n^2-n-6\right)}{n^2\left(n-1\right)^2}.
    \label{eq:Jproperty_3}
\end{align}
\end{subequations}

Similarly, for 1-$2$-SAT$^+$, the the probability that $A_{ia}A_{ja} = 1$ for arbitrary two different $i$ and $j$ is $p_2 = 1/\binom{n}{2}$, and the probability that $J_{ij} = J$ is
\begin{equation}
    P_2\left(J_{ij}=J\right) = \binom{m}{J}p_2^J\left(1-p_2\right)^{m-J},
    \label{eq:Jdist_2}
\end{equation}
which is also size-dependent. The mean and variance are
\begin{subequations}
\begin{align}
    \mathbb{E}_2\left(J_{ij}\right) &= \frac{2m}{n\left(n-1\right)},\\
    {\rm Var}_3\left(J_{ij}\right) &= \frac{2m\left(n^2-n-2\right)}{n^2\left(n-1\right)^2}.
    \label{eq:Jproperty_2}
\end{align}
\end{subequations}


As the ratio between standard deviation to the mean of $J_{ij}$ decreases with the clause-to-variable ratio $m/n$ for fixed $n$, we expect in the limit of large $m/n$, the coefficients $J_{ij}$ approach uniform for all $i,j$. The same applies to $h_i$'s for $H_{C,3^+}$. At the end of the discussion, we want to address the difference between 1-$k$-SAT$^+$ and the well-known Sherrington-Kirkpatrick (SK) model in spin glass with the Hamiltonian $H_{SK}=\sum_{i<j} J_{ij}\sigma_i^z\sigma_j^z$ where $J_{ij}$ is independently sampled from the standard normal distribution $\mathcal{N}(0,1)$~\cite{farhi2019quantum}.

\subsection{Gate-based implementation of QAOA}
\label{app:QAOA_implementation}

To implement the Hamiltonian dynamics in QAOA with a quantum circuit, one can decompose the unitary evolution into parallel Pauli-X and Pauli-Z gates as the following
\begin{align}
    &\exp{-i\gamma_k H_{C,3}}
    \nonumber
    \\
    &=  \exp\left(i\frac{\gamma_k}{8}\sum_i h_i\sigma_i^z\right)\exp\left(-i\frac{\gamma_k}{8}\sum_{i<j} J_{ij}\sigma_i^z\sigma_j^z\right)\nonumber\\ &\exp\left(-i\frac{\gamma_k}{8}\sum_{i<j<\ell} K_{ij\ell}\sigma_i^z\sigma_j^z\sigma_\ell^z\right) 
    \nonumber 
    \\
    & = \prod_i \exp\left(i\frac{\gamma_kh_i}{8}\sigma_i^z\right)\prod_{i<j} \exp\left(-i\frac{\gamma_k J_{ij}}{8}\sigma_i^z\sigma_j^z\right)\nonumber\\
    &\prod_{i<j<\ell}
    \exp\left(-i\frac{\gamma_k K_{ij\ell}}{8}\sigma_i^z\sigma_j^z\right) \label{eq:HC_3sat_2}
\end{align}
for $3$-SAT and it is similar for $H_{C,2}$ by taking $K_{ij\ell}=0$ and a factor of $2$ in the denominator of exponents. For $1$-$3$-SAT$^+$, the problem Hamiltonian layer is
\begin{align}
    &\exp{-i\gamma_k H_{C,3^+}}
    \nonumber
    \\
    &=  \exp\left(-i\frac{\gamma_k}{2}\sum_i h_i\sigma_i^z\right)\exp\left(-i\frac{\gamma_k}{2}\sum_{i<j} J_{ij}\sigma_i^z\sigma_j^z\right) 
    \nonumber 
    \\
    & = \prod_i \exp\left(-i\frac{\gamma_kh_i}{2}\sigma_i^z\right)\prod_{i<j} \exp\left(-i\frac{\gamma_k J_{ij}}{2}\sigma_i^z\sigma_j^z\right) \label{eq:HC_2}.
\end{align}
The case of $H_{C,2^+}$ is similar, with all $h_i$'s equaling zero.
The first and second product in Eq.~\eqref{eq:HC_3sat_2} and ~\eqref{eq:HC_2} correspond to parallel Pauli-Z rotation (RZ) rotation with ZZ interaction (RZZ) gates); \QZ{and the unique third product in Eq.~\eqref{eq:HC_3sat_2} correspond to rotaion with ZZZ interaction (RZZZ gates)}. Similarly, $\exp\left(-i\beta_kH_B\right)$ is implemented by Pauli-X rotation (RX) gates. 


Numerically, we implement the QAOA with Qulacs~\cite{suzuki2020qulacs}, a high-performance quantum computing platform for both Python and \verb!C++!. We employ the 
Broyden–Fletcher–Goldfarb–Shanno (BFGS) algorithm~\cite{broyden1970convergence,fletcher1970new,goldfarb1970family,shanno1970conditioning}, a gradient-based quasi-Newton method implemented in Scipy~\cite{virtanen2020scipy}, to find the optimal parameters $\vec{\gamma}^*,\vec{\beta}^*$. \QZ{The classical optimization stops when either the difference of cost function between steps or gradient norm is smaller than $10^{-6}$.} Our numerical simulations are performed on the Puma HPC from University of Arizona with $50$ cores of AMD Zen2 CPU and $250$GB of RAM.

\subsection{Details of the classical approximate algorithms}
\label{app:MWIS-algorithms}

Variants of greedy algorithms are proposed for approximate MWIS problems~\cite{sakai2003note, kako2005approximation}. Before applying those algorithms for benchmark, we introduce some notations to avoid any confusion. Given a weighted graph $G(Q,E,w)$, where $Q,E,w$ represent the set of vertices, edges and weights of vertices, we use $w(q_i)$ denotes the weight of vertex $q_i$ and $w(S)$ denotes the sum of weight for vertex set $S$. $N(q_i)$ represents the set of vertices that are adjacent to vertex $q_i$ and $N^+(q_i)=N(q_i)\cup\{q_i\}$. We denote the degree of vertex $q$ in graph $G_i$ as $d_{G_i}(q)$.

We briefly summarize the four greedy algorithms that are used for benchmark in this paper, GWMIN, GWMAX, GWMIN2~\cite{sakai2003note} and WG~\cite{kako2005approximation} algorithms. We also list their corresponding guaranteed lower bounds on maximum weight estimation.

\begin{algorithm}
\label{alg:gwmin}
\begin{algorithmic}
{\rm GWMIN}
\State {\rm Begin} $S=\emptyset, i=0, G_i=G$
\While{$Q(G_i) \neq \emptyset$}
\State  {\rm Choose a vertex} $q$ s.t. $q = {\rm argmax}_{u\in Q(G_i)}\frac{w(u)}{d_{G_i}(u)+1}$
\State $S = S\cup\{q\}$; {\rm remove $N_{G_i}^+(q)$ from $G_i$}; $i=i+1$
\EndWhile
\State {\rm Output} $S$
\end{algorithmic}
\end{algorithm}

\begin{algorithm}
\label{alg:gwmax}
\begin{algorithmic}
{\rm GWMAX}
\State {\rm Begin} $S=\emptyset, i=0, G_i=G$
\While{$E(G_i) \neq \emptyset$}
\State  {\rm Choose a vertex} $q$ s.t.
\State $q = {\rm argmin}_{u\in Q(G_i)}\frac{w(u)}{d_{G_i}(u)\left(d_{G_i}(u)+1\right)}$
\State {\rm remove $q$ from $G_i$}; $i=i+1$
\EndWhile
\State {\rm Output} $S=Q(G_i)$
\end{algorithmic}
\end{algorithm}

\begin{algorithm}
\label{alg:gwmin2}
\begin{algorithmic}
{\rm GWMIN2}
\State {\rm Begin} $S=\emptyset, i=0, G_i=G$
\While{$Q(G_i) \neq \emptyset$}
\State  {\rm Choose a vertex} $q$ s.t.
\State $q = {\rm argmax}_{u\in Q(G_i)}\frac{w(u)}{\sum_{q\in N_{G_i}^+(u)}w(q)}$
\State $S = S\cup\{q\}$; {\rm remove $N_{G_i}^+(q)$ from $G_i$}; $i=i+1$
\EndWhile
\State {\rm Output} $S$
\end{algorithmic}
\end{algorithm}

\begin{algorithm}
\label{algorithm:wg}
\begin{algorithmic}
{\rm WG}
\State {\rm Begin} $S=\emptyset, i=0, G_i=G$
\While{$Q(G_i) \neq \emptyset$}
\State  {\rm Choose a vertex} $q$ s.t.
\State $q = {\rm argmin}_{u\in Q(G_i)}\frac{\sum_{q\in N_{G_i}(u)}w(q)}{w(u)}$
\State $S = S\cup\{q\}$; {\rm remove $N_{G_i}^+(q)$ from $G_i$}; $i=i+1$
\EndWhile
\State {\rm Output} $S$
\end{algorithmic}
\end{algorithm}
It is also shown that the lower bounds of approximate ratio for those approximate algorithms are
\begin{subequations}
\begin{align}
r_{\rm GWMIN} &\ge \sum_{q\in Q(G)}\frac{w(q)}{d_G(q)+1}\,,\\
r_{\rm GWMAX} &\ge \sum_{q\in Q(G)}\frac{w(v)}{d_G(q)+1}\,,\\
r_{\rm GWMIN2} &\ge \sum_{q\in Q(G)}\frac{w(q)^2}{\sum_{u \in N_G^+(q)} w(u)}\,,\\
r_{\rm WG} &\ge \frac{W(G)}{\sum_{q\in Q(G)}w(N_G(q))/w(G)+1}.
\end{align}
\label{eq:alg_bound}
\end{subequations}

\begin{figure}
    \centering
    \includegraphics[width=0.45\textwidth]{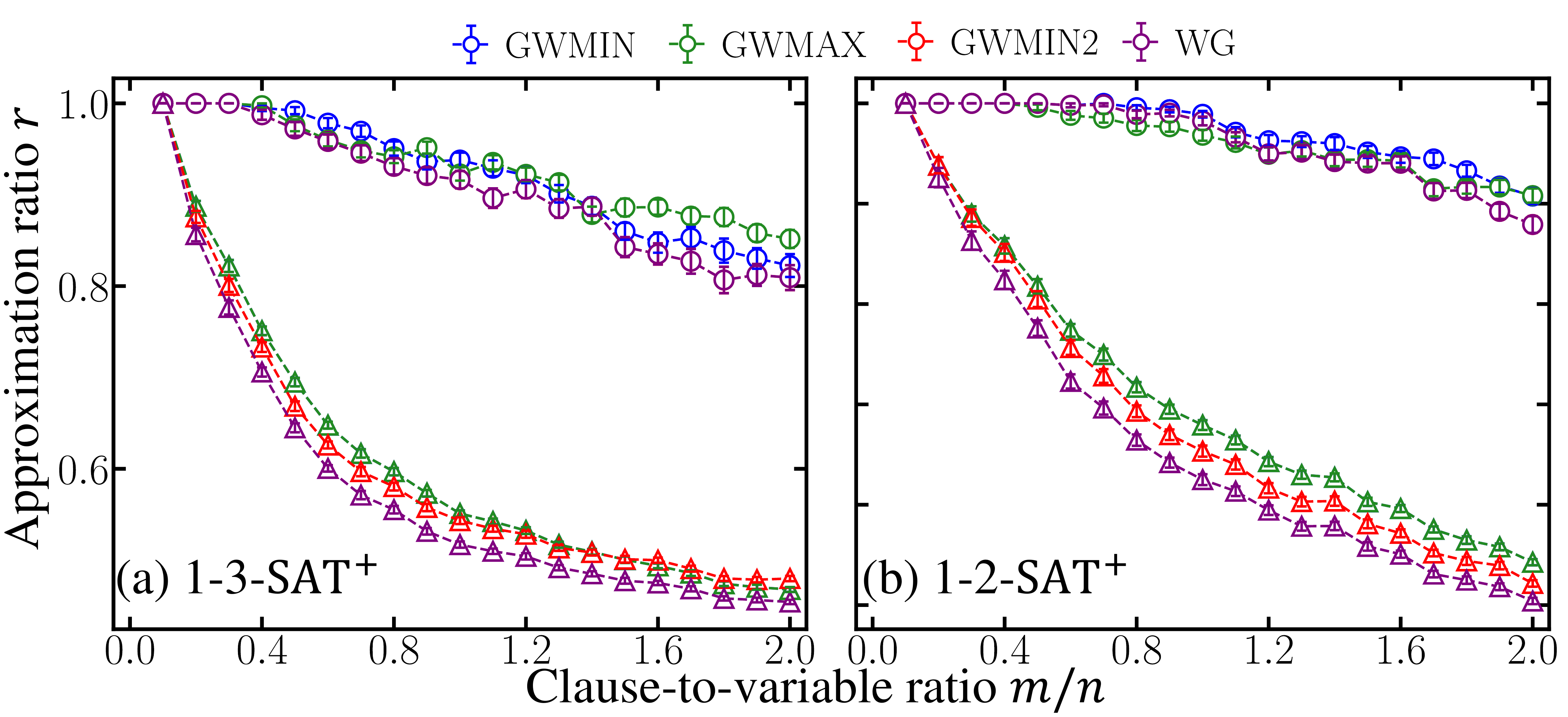}
    \caption{Benchmark on approximate ratio $r$ with MWIS approximate algorithms for instances reduced from (a) 1-$3$-SAT$^+$ and (b) 1-$2$-SAT$^+$. Circles represent approximate ratio $r$ of different algorithms and triangles represent the corresponding lower bound. All results are fixed with $n=10$ variables.}
    \label{fig:mis_classical}
\end{figure}

To obtain benchmarks, we reduce the 1-$k$-SAT$^+$ instances to MWIS instances with igraph~\cite{csardi2006igraph} and apply the approximate algorithms introduced above to obtain the approximation ratio results in Fig.~\ref{fig:mis_classical}. As all the four algorithms are variants of greedy algorithms, their performances are similar. The lower bounds of those algorithms in Eq.~\eqref{eq:alg_bound} are also plotted in Fig.~\ref{fig:mis_classical}(a)(b) as a reference. The final benchmark presented in the main paper are obtained from the best approximation ratio among the four algorithms for every fixed clause-to-variable ratio $m/n$ separately.

\subsubsection{On approximation ratio}
Here we provide a brief summary of known facts of the approximation ratio of problems related to the 1-$k$-SAT$^+$, as there are not much known results for 1-$k$-SAT$^+$ itself. Note that these results do not carry over to the 1-$k$-SAT$^+$, as we explain below.

One can reduce a 1-$k$-SAT$^+$ instance to a $k$-SAT instance. 
There is a simple polynomial-time algorithm that provides a $(1-1/{2^k})$ approximation ratio for Max-$k$-SAT (in this paper we always mean exact $k$ variables in each clause). This is also equal to the expected approximation ratio of a random assignment.
\QZ{Ref.~\cite{interian2004approximation} shows that the lower bound of approximation ratio for polynomial-time algorithm in solving Max-$3$-SAT to be $r\ge 0.95$.} Ref.~\cite{haastad2001some} also shows that it is NP-hard to approximate Max-$2$-SAT with any approximation ratio above $21/22\simeq 0.955$. However, it is important to note that the above are worst case results and does not directly apply to 1-$k$-SAT$^+$ due to the reduction.

There are also results on approximate algorithms guaranteeing certain approximate ratios for random instances of Max-$k$-SAT. However, as instances generated by reducing the random instances of 1-$k$-SAT$^+$ to $k$-SAT are by no means random, these results also do not apply to random instances of Max-1-$k$-SAT$^+$.  




\subsection{More details on QAOA performance}
\label{app:QAOA_details}


\begin{figure}[t]
    \centering
    \includegraphics[width=0.45\textwidth]{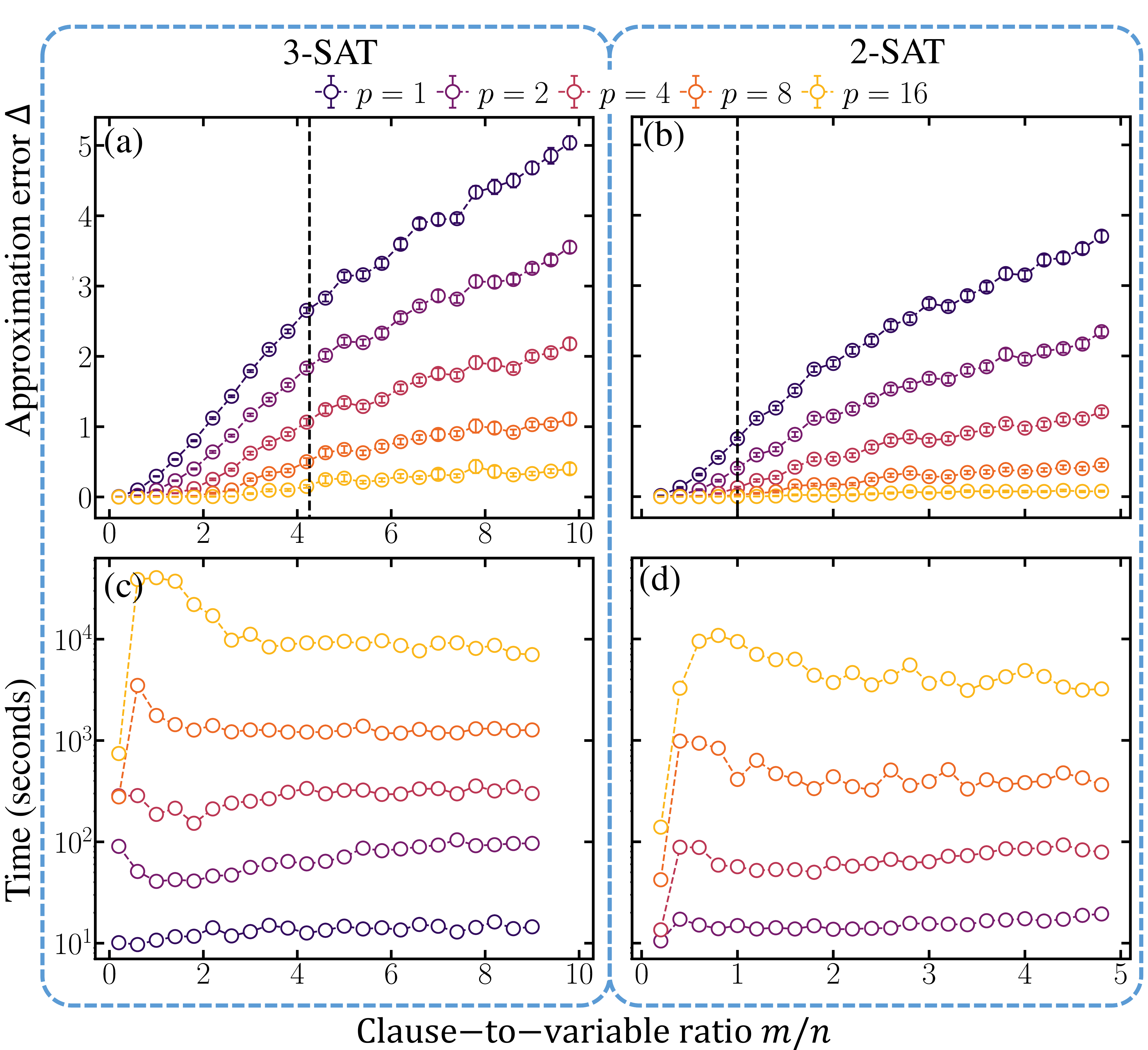}
    \caption{(a)(b) Approximation error $\Delta$, and (c)(d) CPU time (in seconds) for a $p$-layer QAOA versus clause-to-variable ratio $m/n$ to solve $3$-SAT (left) and $2$-SAT (right) with $n=10$ variables. Black dashed line in (a), (b) represents the critical point of SAT-UNSAT transition. \label{fig:QAOA_kSAT_deltaE}}
    \centering
    \includegraphics[width=0.45\textwidth]{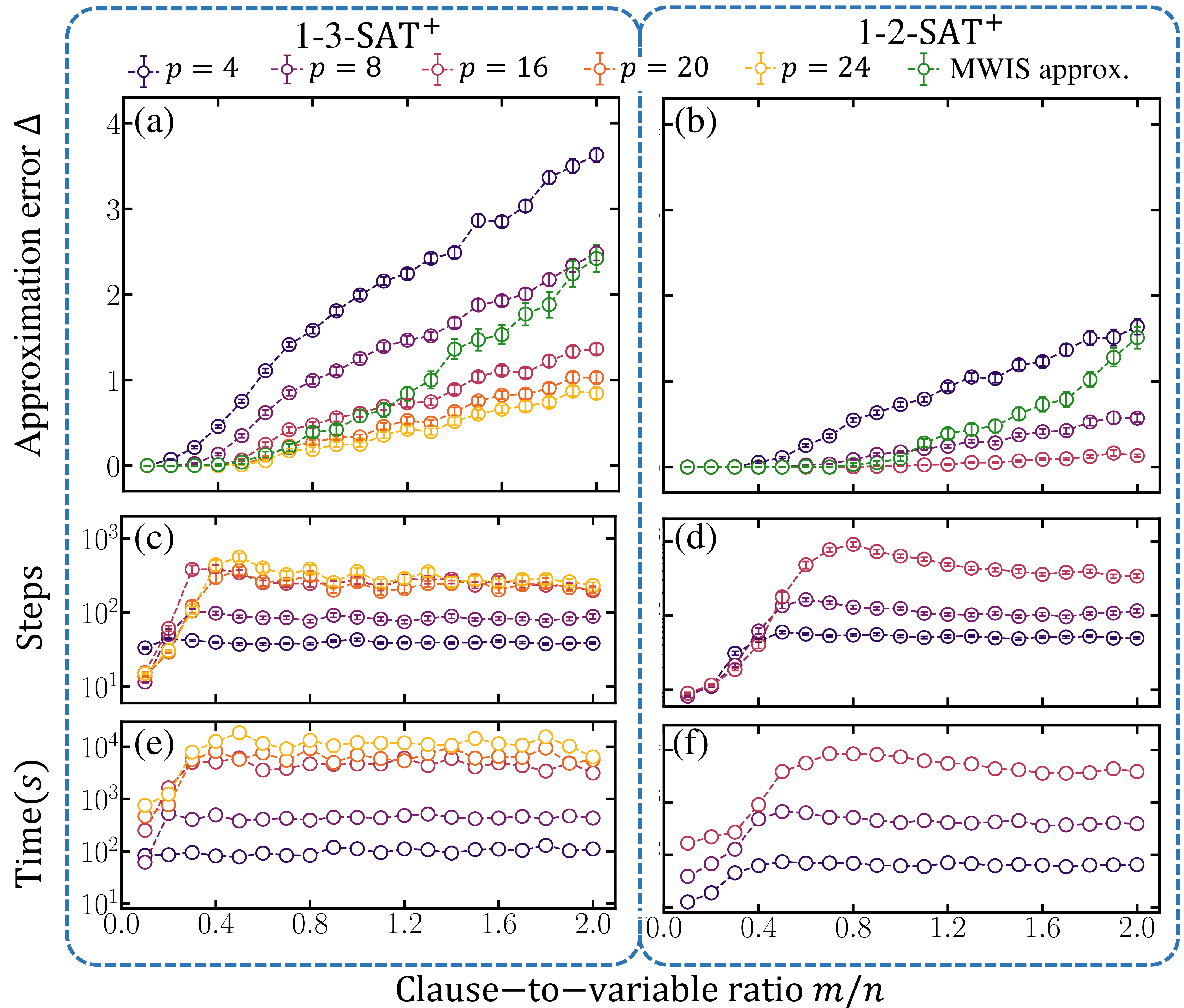}
    \caption{(a)(b) Approximation error $\Delta$, (c)(d) optimization steps and (e)(f) CPU time (in seconds) for a $p$-layer QAOA versus clause-to-variable ratio $m/n$ to solve 1-$3$-SAT$^+$ (left) and 1-$2$-SAT$^+$ (right) with $n=10$ variables. Green dots in (a)(b) represent classical approximate solutions obtained by reducing to the MWIS problem.
    \label{fig:QAOA_deltaE}}
\end{figure}

\subsubsection{QAOA performance and computing time}
\QZ{In the main text, we have provided the approximation ratio results in Fig.~\ref{fig:qaoa_kSAT_ratio} and Fig.~\ref{fig:qaoa_ratio}. Here we also plot the absolute error $\Delta$ (number of additional violated clauses) for $k$-SAT and $1$-$k$-SAT$^+$ in Fig.~\ref{fig:QAOA_kSAT_deltaE}, Fig.~\ref{fig:QAOA_deltaE}(a) and (b).
The optimization steps towards the optimal solution and the computing time are also shown in Fig.~\ref{fig:QAOA_kSAT_deltaE}(c)(d) and Fig.~\ref{fig:QAOA_deltaE}(c)-(f).}

We point out that for 1-$3$-SAT$^+$ the ground energy $E_0$ of $H_{C,3^+}$ could be different from the number of violated clauses, because the energy cost of a violated clause can be different among possible spin configurations; While for 1-$2$-SAT$^+$ there is no such difference. 
This difference means that
despite $H_{C,3^+}$ is a well-accepted Hamitlonian for 1-$3$-SAT$^+$ problem~\cite{choi2010adiabatic,zhuang2014increase}, there may be other better options.

\begin{figure}[t]
    \centering
    \includegraphics[width=0.45\textwidth]{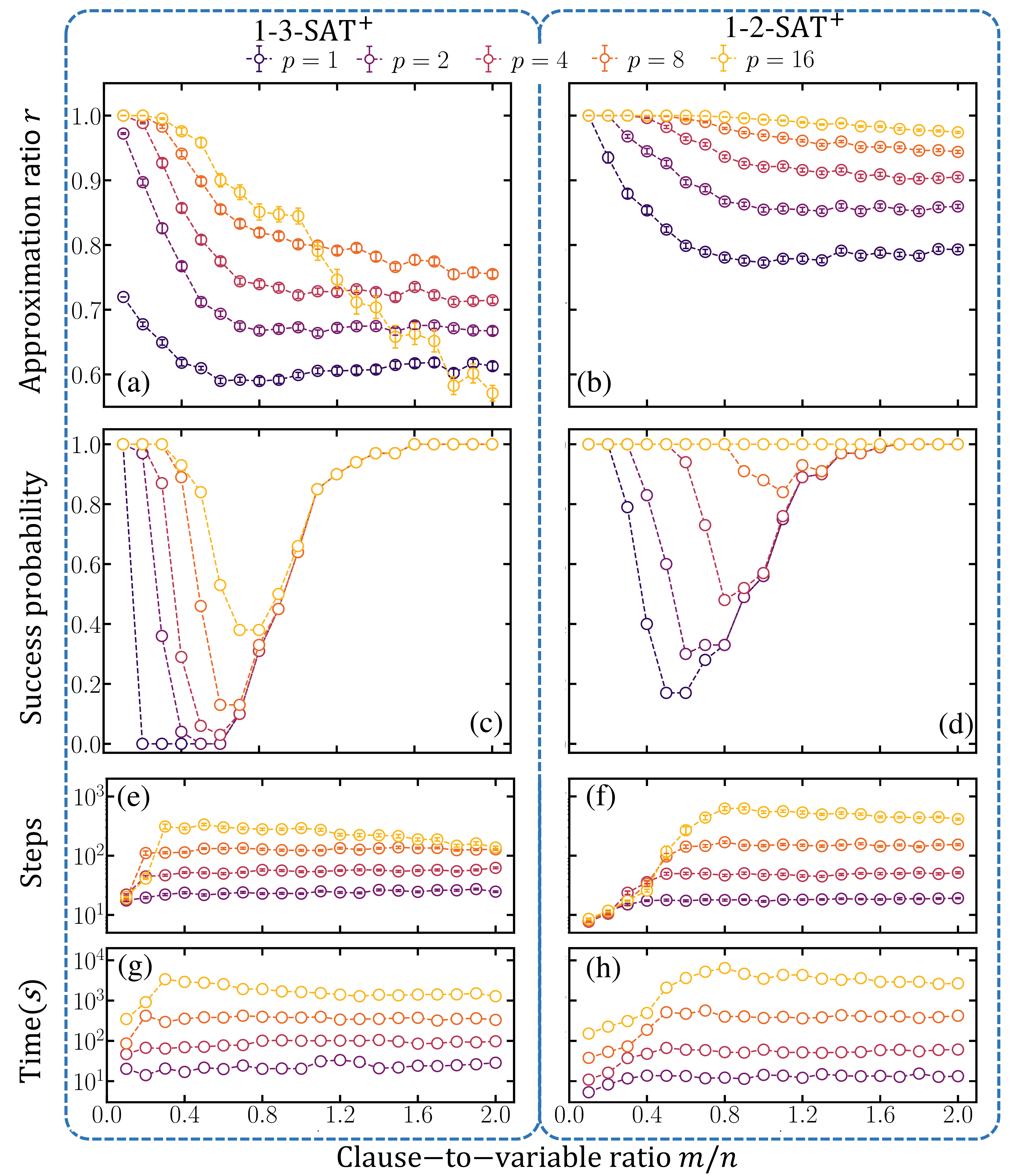}
    \caption{(a)(b) Approximation ratio $r$, (c)(d) success probability, (e)(f) optimization steps and (g)(h) CPU computing time (in seconds) for a $p$-layer QAOA versus clause-to-variable ratio $m/n$ to solve 1-$3$-SAT$^+$ (left) and 1-$2$-SAT$^+$ (right) with $n=10$ variables. All results are obtained with the simple random initialization strategy.}
    \label{fig:QAOA_rinit}
\end{figure}

\subsubsection{Initialization strategy of QAOA}
For a $p$-layer QAOA, the pre-optimization strategy initializes the first $p^\prime$ layers ($1\le p^\prime< p$) by an optimized $p^\prime$-layer QAOA~\footnote{In practice, the number of layers $p^\prime$ is chosen to be comparable to $p$ to obtain better performance.}, and sample the rest of the parameters $\{\gamma_k\}_{k=p^\prime+1}^{p},\{\beta_k\}_{k=p^\prime+1}^{p}$ randomly uniformly in $[0,\epsilon]$. With the initialization, further training gives the optimal parameters. Here we choose $\epsilon=0.1$ to take advantage of the $p^\prime$-layer QAOA results without being trapped in local minima.

Now we compare the pre-optimization strategy to a simple random initialization strategy, where all initial parameters $\{\gamma_k\}_{k=1}^p,\{\beta_k\}_{k=1}^p$ are randomly sampled uniformly in $[0,\pi]$. We show the performance on Max-1-$k$-SAT$^+$ in Fig.~\ref{fig:QAOA_rinit}(a)(b). Compared to the pre-optimization strategy in Fig.~\ref{fig:qaoa_ratio}(a)(b) of the main paper, the simple random initialization strategy leads to a worse approximation ratio. In particular, as the number of layers $p$ increases, more parameters need to be optimized and the performance of the random initialization strategy
gets much worse than the pre-optimization strategy, especially for $k=3$ with a more complex $H_{C,k}$. As for the decision version, the performance with the simple random initialization strategy (shown in Fig.~\ref{fig:QAOA_rinit}(c),(d)) is still worse than the performance of the pre-optimization strategy in Fig.~\ref{fig:qaoa_ratio}(c),(d) of the main paper. In Fig.~\ref{fig:QAOA_rinit}(e)-(h), we plot the steps and computing time for the random strategy; Comparing with Fig.~\ref{fig:QAOA_deltaE}(c)-(f), we see the computation cost is at the same order of magnitude with the pre-optimization strategy.

\begin{figure}[t]
    \centering
    \includegraphics[width=0.45\textwidth]{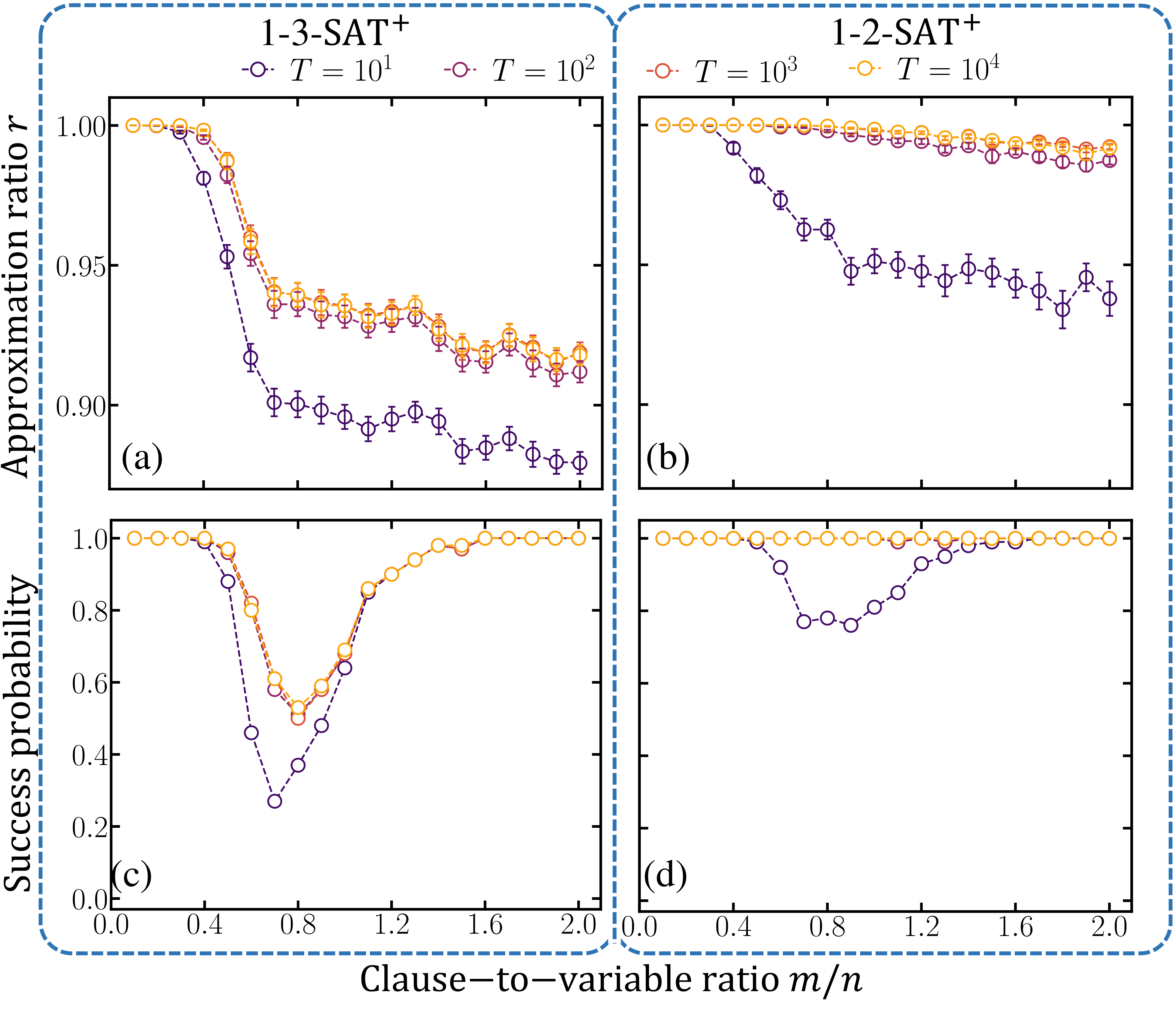}
    \caption{(a)(b) Approximation ratio $r$ and (c)(d) success probability of a $p=16$-layer QAOA versus clause-to-variable ratio $m/n$ for 1-$3$-SAT$^+$ (left) and 1-$2$-SAT$^+$ (right) with $n=10$ variables. Dots from dark to light represent the cutoff of optimization steps at $T=10, 100, 1000, 10000$. \label{fig:QAOA_Tcut}
    }
\end{figure}

\subsubsection{Performance with limitations}
With the existence of barren plateaus at a large depth, finding the best optimal parameters $\vec{\gamma}^*,\vec{\beta}^*$ in QAOA consumes a large amount of computing resources for large problem instances. At the same time, the accumulation of errors and noise in near-term devices prohibits a large quantum system to be stable for a long time~\cite{preskill2018quantum}. Therefore, we consider the sub-optimal performance of QAOA under a cutoff $T$ on the optimization steps, shown in Fig.~\ref{fig:QAOA_Tcut}.

It turns out that for the $n=10$ qubit system with $p=16$ layers, the performance saturates quickly at only around a hundred steps in most parameter regions. In particular, for the decision version of the problem (Fig.~\ref{fig:QAOA_Tcut} (c)(d)), the easy problems away from the transition only takes a few optimization steps to solve. The only exception is the Max-1-$k$-SAT$^+$ problem at large $m/n$, where around a thousand steps are necessary. This is due to the optimization versions of the problem being harder at large $m/n$ ratios, recovering the reachability deficits~\cite{akshay2020reachability}. Overall, for a given $p$, it is efficient to take a limited number of optimization steps in practical implementations to get a balance between accuracy and resource consumption.


%

\end{document}